\documentclass[useAMS,usenatbib,usegraphicx]{mn2e}
\usepackage{natbib}
\usepackage{color}
\usepackage{hyperref} 
\hypersetup{colorlinks=true,citecolor=blue}
\usepackage{times}
\usepackage[T1]{fontenc}
\usepackage{aecompl}
\def \bea {\begin{eqnarray}}
\def \ena {\end{eqnarray}}               
\def \bee {\begin{equation}}
\def \ene {\end{equation}}
\def    \simlt  {\lower.5ex\hbox{$\; \buildrel < \over \sim \;$}}
\def    \simgt  {\lower.5ex\hbox{$\; \buildrel > \over \sim \;$}}

\def	\ltsim	{\simlt}

\newcommand     \mum    {\,\mu{\rm m}}  

\def	\cm		{\,{\rm {cm}}}

\def	\B		{{\rm B}}

\def	\erg		{\,{\rm {ergs}}}

\def    \exp 		{\,{\rm {exp}}}
\def	\g		{\,{\rm g}}

\def	\K		{\,{\rm K}}

\def	\pc		{\,{\rm {pc}}}

\def	\s		{\,{\rm s}}

\def    \ln  		{\,{\rm {ln}}}
\def    \yr  		{\,{\rm {yr}}}

\def	\H		{\rm H}

\def	\xhat		{\hat{\bf x}}
\def	\yhat		{\hat{\bf y}}
\def	\zhat		{\hat{\bf z}}
\def	\Xhat		{\hat{\bf X}}
\def	\Yhat		{\hat{\bf Y}}
\def	\Zhat		{\hat{\bf Z}}
\def	\V		{\rm V}

\def	\ehat		{\hat{\bf e}}

\def    \Bv     	{\bf  B}

\def    \rv     	{{\bf  r}}
\def    \kv     	{\bf  k}

\def	\ba		{{\bf a}}
\def	\be		{{\bf e}}

\def	\bB		{{\bf B}}

\def 	\bE		{{\bf E}}

\def	\bJ		{{\bf J}}
\def	\bk		{\kv}

\def   	\bQ  		{{\bf Q}}

\def	\bv		{{\bf v}}

\def	\gas		{\rm {gas}}
\def	\th		{\rm {th}}

\def	\d		{\rm d}
\def	\rad		{\rm {rad}}

\def    \ext    	{\rm {ext}}
\def    \pol    	{\rm {pol}}

\def	\kin		{\rm {kin}}

\def    \coll        	{\rm {coll}}

\def	\ISRF		{\rm {ISRF}}

\def    \RAT		{\rm {RAT}}

\def	\hi		{\rm {highJ}}
\def	\lo		{\rm {lowJ}}

\def	\ali		{\rm {ali}}

\font\mib=cmmib10
\def\balpha{\hbox{\mib\char"0B}}
\def\bbeta{\hbox{\mib\char"0C}}
\def\bgamma{\hbox{\mib\char"0D}}
\def\bGamma{\hbox{\mib\char"00}}

\def\bdelta{\hbox{\mib\char"0E}}
\def\bepsilon{\hbox{\mib\char"0F}}
\def\bzeta{\hbox{\mib\char"10}}
\def\boldeta{\hbox{\mib\char"11}}
\def\btheta{\hbox{\mib\char"12}}
\def\biota{\hbox{\mib\char"13}}
\def\bkappa{\hbox{\mib\char"14}}
\def\blambda{\hbox{\mib\char"15}}
\def\bmu{\hbox{\mib\char"16}}
\def\bnu{\hbox{\mib\char"17}}
\def\bxi{\hbox{\mib\char"18}}
\def\bpi{\hbox{\mib\char"19}}
\def\brho{\hbox{\mib\char"1A}}
\def\bsigma{\hbox{\mib\char"1B}}
\def\btau{\hbox{\mib\char"1C}}
\def\bupsilon{\hbox{\mib\char"1D}}
\def\bphi{\hbox{\mib\char"1E}}
\def\bchi{\hbox{\mib\char"1F}}
\def\bpsi{\hbox{\mib\char"20}}
\def\bomega{\hbox{\mib\char"21}}

\title{Modeling Grain Alignment by Radiative Torques and Hydrogen Formation Torques in Reflection Nebula}
\author[Thiem Hoang, A. Lazarian, and B-G Andersson]{Thiem Hoang $^{1,2}$\thanks{E-mail: hoang@cita.utoronto.ca}, A. Lazarian$^3$, and B-G Andersson$^4$\\
$^1$ Institut f$\ddot{\rm u}$r Theoretische Physik, Lehrstuhl IV: Weltraum- und 
Astrophysik, Ruhr-Universit$\ddot{\rm a}$t Bochum, 44780 Bochum, Germany,\\
$^2$ Canadian Institute for Theoretical Astrophysics, University of Toronto, 60 St. George Street, Toronto, ON M5S 3H8, Canada,\\
$^3$ Department of Astronomy, University of Wisconsin-Madison, Madison, WI 53705, USA,\\
$^4$ SOFIA Science Center, Universities Space Research Association, NASA Ames Research Center, M.S. N232-12 Moffett Field, CA 94035, USA}

\begin{document}
\maketitle

\begin{abstract}
Reflection nebulae--dense cores--illuminated by surrounding stars offer a unique opportunity to directly test our quantitative model of grain alignment based on radiative torques (RATs) and to explore new effects arising from additional torques. In this paper, we first perform detailed modeling of grain alignment by RATs for the IC 63 reflection nebula illuminated both by a nearby $\gamma$ Cas star and the diffuse interstellar radiation field. We calculate linear polarization $p_{\lambda}$ of background stars by radiatively aligned grains and explore the variation of fractional polarization ($p_{\lambda}/A_V$) with visual extinction $A_{V}$ across the cloud. Our results show that the variation of $p_{V}/A_{V}$ versus $A_{V}$ from the dayside of IC 63 to its center can be represented by a power-law ($p_{V}/A_{V}\propto A_{V}^{\eta}$) with different slopes depending on $A_{V}$. We find a shallow slope $\eta \sim- 0.1$ for $A_{V}< 3$ and a very steep slope $\eta\sim -2$ for $A_{V}> 4$. We then consider the effects of additional torques due to $\H_{2}$ formation and model grain alignment by joint action of RATs and H$_2$ torques. We find that $p_{V}/A_{V}$ tends to increase with an increasing magnitude of H$_{2}$ torques. In particular, the theoretical predictions obtained for $p_{V}/A_{V}$ and peak wavelength $\lambda_{\max}$ in this case show an improved agreement with the observational data. Our results reinforce the predictive power of the RAT alignment mechanism in a broad range of environmental conditions and show the effect of pinwheel torques in environments with efficient H$_2$ formation. Physical parameters involved in H$_2$ formation may be constrained using detailed modeling of grain alignment combined with observational data. In addition, we discuss implications of our modeling for interpreting latest observational data by {\it Planck} and other ground-based instruments.
\end{abstract}

\begin{keywords}
magnetic fields- polarization- dust, extinction
\end{keywords}

\section{Introduction}\label{sec:intro}
Immediately after the discovery of polarization of light from distant stars, more than 60 years ago by \cite{Hall:1949p5890} and \cite{Hiltner:1949p5851}, the polarization was attributed to differential extinction by nonspherical dust grains aligned with interstellar magnetic fields. This alignment of grains opened a new window into studying the magnetic fields, including the magnetic field strength through starlight polarization (\citealt{1951ApJ...114..206D}; \citealt{1953ApJ...118..113C}) and polarized thermal dust emission (\citealt{Hildebrand:1988p2566}), in various astrophysical environments. Moreover, polarized thermal emission from aligned grains is a significant Galactic foreground source contaminating cosmic microwave background experiments (\citealt{2009AIPC.1141..222D}; \citealt{2014arXiv1409.5738P}). However, it is only recently that grain alignment theory has become quantitative and predictive, which allows for realistic modeling of dust polarization and direct comparison with observations.

The problem of grain alignment has proven to be one of the longest standing problems in astrophysics. Over the last 60 years, a number of grain alignment mechanisms have been proposed and quantified (see \citealt{2007JQSRT.106..225L} for a review).  Some substantial extensions or modifications were suggested to the initial paradigm of grain alignment based on the \cite{1951ApJ...114..206D} paramagnetic relaxation theory.  However, an alternative alignment paradigm, based on radiative torques (RATs), has now become the favored mechanism to explain grain alignment. This mechanism was initially proposed by \cite{1976Ap&SS..43..291D}, but was mostly ignored at the time of its introduction due to the limited ability to generate quantitative theoretical predictions. \cite{1996ApJ...470..551D} and \cite{1997ApJ...480..633D} reinvigorated the study of the RAT mechanism by developing a numerical method based on discrete dipole approximation to compute RATs for several irregular grain shapes. The strength of the torques obtained made it impossible to ignore them, but questions about basic properties (e.g., direction, dependence on grain size and shape) of the alignment for grains of different shapes as well as degree of alignment remained. 

The quantitative study of RAT alignment was initiated in a series of papers, by our group, starting with \cite{2007MNRAS.378..910L} (henceforth LH07) where an analytical model of RAT alignment was introduced. The analytical model was the basis for further theoretical studies in (\citealt{Lazarian:2007p2442}; \citealt{Lazarian:2008fw}, \citealt{{2009ApJ...695.1457H},{2009ApJ...697.1316H}}; see also reviews in \citealt{2007JQSRT.106..225L}; Lazarian, Andersson, \& Hoang 2015; Andersson, Lazarian, \& Vaillancourt 2015).

This work clarified why the grain alignment occurs with the long axes perpendicular to the magnetic field (as required by observations), even though the magnetic field provides only the axis of alignment, which earlier seemed to permit both the alignment perpendicular and parallel to the magnetic field.  These studies opened a way for quantitatively predicting the grain alignment for a variety of astrophysical situations (see \citealt{2014MNRAS.438..680H}). 

The basic requirement for achieving grain alignment in the RAT paradigm is that grains with a net helicity are embedded in a magnetic field and exposed to anisotropic radiation with a wavelength less than the grain diameter. The radiation field also must have sufficient energy density. The grain's helicity causes a difference in the scattering cross section to the left- and right-hand circular polarization components of the radiation field, imparting a torque on the grain.  As the grain gets magnetized through the Barnett effect (\citealt{Barnett:1915p6353}; \citealt{1976Ap&SS..43..291D}), it Larmor precesses around the magnetic field lines. The continued action of the radiative torques on the spinning grain can then lead to alignment of grain angular momentum $\bJ$ with the magnetic field $\Bv$.

In particular, we found that RATs tend to align grains at attractor points with low magnitude of angular momentum (i.e., $J \sim J_{\th}$ with $J_{\th}$ being the thermal angular momentum, hereafter low-$J$ attractor points), and/or attractor points with high angular momentum (i.e., $J> J_{\th}$, hereafter high-$J$ attractor points). The high-$J$ attractor points mostly correspond to the perfect alignment of $\bJ$ with $\Bv$ (i.e., $\cos\beta=\pm 1$ with $\beta$ the alignment angle made by $\bJ$ and $\Bv$), while the low-$J$ attractor points occur at $\cos\beta=\pm 1$ or in its vicinity. The existence of high-$J$ attractor points depends on the gas density and temperature, the angle between the radiation field anisotropy and the magnetic field, the grain size, shape and composition, the radiation field. Within our analytical model (AMO) of RATs, the four last parameters can be combined into a single parameter $q^{\rm max}$, which is the ratio of the RAT efficiency component parallel to the radiation field anisotropy direction to the component perpendicular to it (see Appendix \ref{apdx:Jmax-psi}). We expect a reduction of the order of 20-30 percent in the degree of alignment when the alignment only happens with the low-$J$ attractor points. Both superparamagnetic inclusions and H$_2$ formation torques had been considered as ways to increase the efficiency of paramagnetic alignment (\citealt{Jones:1967p2924}; \citealt{1979ApJ...231..404P}; \citealt{Spitzer:1979p2708}). It is interesting that both processes were found to be important for the RAT alignment.

For instance, in \cite{Lazarian:2008fw}, we found that the existence of strongly magnetic inclusions within the grain creates high-$J$ attractor points and thus perhaps makes the RAT alignment eventually perfect. In that sense, the inclusion of iron-rich clusters into dust grains first increases the fraction of grains aligned with high-$J$ attractor points. Then gas bombardment can accidentally move some grains from the low-$J$ to high-$J$ attractor point and increases the alignment in this way. The dynamics of grain alignment is pretty complex with grains undergoing thermal flipping (\citealt{1999ApJ...516L..37L}).\footnote{The picture of thermal flipping was challenged by \cite{Weingartner:2009p5709} who found that the grain does not experience thermal flipping as a result of internal relaxation, instead, it tends to be frozen at the separatrix (i.e., when the grain symmetry axis becomes perpendicular to the angular momentum). \cite{2009ApJ...695.1457H}, however, showed that the excitation by gas bombardment and H$_2$ formation can prevent the grain from being frozen at the separatrix. Thus, dust grains do flip.}

In addition to RATs, other systematic torques can act on grains. \cite{1979ApJ...231..404P} proposed three surface processes that produce systematic torques, including H$_{2}$ formation on random active sites, the variation of photoelectric yield and of sticking coefficient of gas atoms on the grain surface. These torques are frequently referred to as Purcell (or pinwheel) torques, and the torques due to H$_{2}$ formation are expected to be dominant. \cite{2009ApJ...695.1457H} suggested infrared emission from irregular grains as another mechanism producing pinwheel torques. The Purcell torques together with paramagnetic relaxation were at one time thought to be the major mechanism leading to the alignment of $\bJ$ with $\Bv$ \citep{1979ApJ...231..404P}.  However, as shown by LH07, in the presence of RATs, the alignment arising from paramagnetic relaxation for ordinary paramagnetic grains is negligible compared to that arising from RATs.

Because the Purcell torques are fixed within the grain body, their efficiency in aligning the grains was found to decrease when the grains wobble rapidly and becomes negligible when the grains undergo rapidly thermal flipping (\citealt{1999ApJ...516L..37L}). Nevertheless, in the framework of RAT alignment, the Purcell torques were found to enhance the degree of grain alignment through two processes (\citealt{2009ApJ...695.1457H}). Firstly, when the grains are radiatively aligned with high-$J$ attractor points, the presence of H$_{2}$ torques can enhance the degree of alignment by increasing the angular momentum of the high-$J$ attractor point, thus, driving some smaller grains to suprathermal rotation (i.e. with rotation speeds well above the thermal energy of the environment). Secondly, for low-$J$ attractor points, the H$_{2}$ torques can contribute to drive some grains that have sufficiently slow flipping (depending on their size) to suprathermal rotation, i.e., creating new high-$J$ attractor points from low-$J$ attractor points. The latter corresponds to an increased fraction of grains aligned at high-$J$ attractor points compared to the case without H$_{2}$ torques. 

Grain alignment by RATs in molecular clouds has been studied extensively (\citealt{2005ApJ...631..361C}; \citealt{2007ApJ...663.1055B}; \citealt{2008ApJ...674..304W}; \citealt{2009A&A...502..833P}). These studies dealt with the alignment of grains by the attenuated diffuse interstellar radiation field (ISRF). The observational data from \cite{2008ApJ...674..304W} show that the fractional polarization can be fitted with a power-law, $p_{K}/\tau_{K}\propto A_{V}^{-0.52\pm 0.07}$, where $p_{K}$ and $\tau_{K}$ are the polarization and optical depth measured in the K band. A simple one-dimensional modeling of RAT alignment for a dense, uniform starless cloud in Whittet et al. shows that the fractional polarization first decreases slowly with $A_{V}$ (with a slope shallower than the best-fit one) and then declines rapidly as $A_{V}^{-1}$ for $A_{V}> 3$. Such a steep decline of $p_{K}/\tau_{K}$ arises from a significant decrease in grain alignment due to the attenuation of ISRF toward the cloud center (see \citealt{2008ApJ...674..304W}). In the presence of magnetic field turbulence, the fractional polarization for sightlines above the decorrelation length of magnetic fields is considerably reduced, which is expected to produce a power-law $A_{V}^{-0.5}$ for moderate $A_{V}$ \citep{1992ApJ...389..602J}, consistent with the best-fit slope in \cite{2008ApJ...674..304W}. However, only the wandering of magnetic field lines seems to be insufficient to reproduce steep slopes observed in very dense regions (i.e., large $A_{V}$) of starless cores, and the loss of grain alignment predicted by the RAT alignment theory can successfully reproduce the observations (\citealt{Jones:2014vq}; \citealt{2014A&A...569L...1A}). The importance of RAT alignment has also been shown for special (accretion disks, zodiacal cloud) and highly dynamic environments (e.g., cometary coma; see \citealt{2014MNRAS.438..680H} for more details). 

Observational evidences for RAT alignment are numerous and increasingly available (\citealt{2007ApJ...665..369A}; \citealt{2008ApJ...674..304W}; \citealt{2010ApJ...720.1045A}; \citealt{2011PASJ...63L..43M}; \citealt{2011A&A...534A..19A}). In particular, fundamental features of RAT alignment such as the dependence of alignment on anisotropy direction of radiation, have been tested and confirmed by observations (see \citealt{2010ApJ...720.1045A}; \citealt{2011A&A...534A..19A}). However, evidence of enhancement of grain alignment by pinwheel torques has not been reported, until recently.

Recent polarization observations of background stars behind the reflection nebula IC~63 by \cite{2013ApJ...775...84A} show an enhancement in the polarization for those located behind some regions with the strongest H$_2$ fluorescence intensity and an unusually steep dependence of the fractional polarization on $A_{V}$ ($p_{V}/A_{V}\propto A_{V}^{-1.1\pm 0.1}$).  Direct photodissociation of an H$_{2}$ molecule requires a photon of energy $E>14.7$eV, which is beyond the Lyman limit in the ISM. Thus, the destruction of the H$_{2}$ molecule takes place via a two-step process. First the H$_2$ molecule is excited to an upper electronic state by the $\lambda < 1108\AA$ photon, and then it relaxes to the ground state from the excited state. If, after relaxation, the molecule ends up in a vibrational state with $v>14$ , then it dissociates. Since the vibrational state population resulting from the electronic relaxation is determined by quantum mechanics, the fluorescence emission rate is directly proportional to the destruction rate of the molecules. If, as discussed in Andersson et al, the chemical timescale of the gas is much shorter than the macroscopic evolution timescales, then the destruction and reformation of H$_2$ molecules will be in a state of detailed balance.  This implies that the H$_2$ fluorescence intensity can be used as a tracer of the local H$_2$ formation rate.  

Andersson et al. suggested the effect of additional torques from H$_{2}$ formations as a cause for the enhancement of $p_{V}/A_{V}$ of the stars probing the high fluorescence regions. This paper is intended to present a detailed, ab-initio model of grain alignment by both RATs arising from stellar radiation of $\gamma$ Cas as well as the attenuated ISRF and H$_{2}$ torques for the IC 63 nebula. Our results will be compared directly with the observational data, aiming to elucidate the role of RATs as well as H$_{2}$ torques on grain alignment. 

Furthermore, given the moderate total column density of IC~63 (not deep enough to fully exclude the radiation field as discussed in \citealt{2008ApJ...674..304W}), the steep decline of $p_{V}/A_{V}$ seen in IC~63 cannot be explained by a low opacity RAT model with constant grain alignment and grain randomization.  The observed slope is steeper than the standard predictions by RAT alignment for a starless cloud core in which grains are aligned by the attenuated ISRF.  This study also seeks to resolve this quandary. 

The present paper is organized as follows. In \S \ref{sec:alignmech}, we summarize the rotational damping processes and their characteristic timescales. A description of grain alignment by RATs and $\H_{2}$ pinwheel torques is presented in \S \ref{sec21}. In \S \ref{sec:IC63} we describe a general method to model grain alignment by RATs and calculate linear polarization due to aligned grains. Principal results and comparison with observational data are presented in \S \ref{sec:results}. Discussion and summary are presented in \S \ref{sec:dis} and \ref{sec:sum}, respectively.

\section{Rotational Damping}\label{sec:alignmech}
The rotational damping of interstellar grains mainly arises from collisions with gas atoms and emission of infrared photons (\citealt{Purcell:1971p2747}; \citealt{1993ApJ...418..287R}).\footnote{Here we distinguish interstellar grains from polycyclic aromatic hydrocarbons (PAHs) with size less than $100$\AA. For PAHs, the damping by additional processes, such as electric dipole emission, plasma drag, and ion collisions, can be important (\citealt{1998ApJ...508..157D}; \citealt{Hoang:2010jy}).} Below, their characteristic timescales are provided for reference.
\subsection{Dust-Gas Collisions}

Collisions of a grain with gas atoms consist of elastic collisions and sticking, inelastic collisions. In the latter regime, gas atoms temporally stick to the grain surface followed by their evaporation. For elastic collisions and axisymmetric grain shape, the integration of all collisional torques over the grain surface tends to zero. In the grain frame of reference, the mean torque arising from the sticking collisions for the axisymmetric grain rotating around its symmetry axis also tends to zero when averaged over the grain revolving surface, but the evaporation induces a non-zero mean torque parallel to the rotation axis (see \citealt{1993ApJ...418..287R}; \citealt{Lazarian:1997p5348}).

To facilitate numerical estimates, we consider oblate spheroidal grains with moments of inertia $I_{1}>I_{2}=I_{3}$ along the grain's principal axes $\hat{\ba}_{1}$, $\hat{\ba}_{2}$ and $\hat{\ba}_{3}$. Let $I_{\|}=I_{1}$ and $I_{\perp}=I_{2}=I_{3}$. They take the following forms: 
\bea
I_{\|}=\frac{2}{5}Ma_{2}^{2}=\frac{8\pi}{15}\rho a_{1}a_{2}^{4},\\
I_{\perp}=\frac{4\pi}{15}\rho a_{1}a_{2}^{2}\left(a_{1}^{2}+a_{2}^{2}\right),
\ena
where $a_{1}$ and $a_{2}$ are the lengths of the semi-minor and semi-major axes of the oblate spheroid with inverse axial ratio $s=a_{1}/a_{2}<1$, and $\rho$ is the mass density of grain material.

For the sake of consistency, we use the {\it effective} grain size $a$, which is usually defined as the radius of a sphere of equivalent volume as the following:
\bea
a=\left(\frac{3}{4\pi} (4\pi/3) a_{1}a_{2}^{2}\right)^{1/3}=a_{2}s^{1/3}.\label{eq:aeff}
\ena

The decrease of grain angular momentum due to the dust-gas collisions is governed by
\bea
\frac{\langle \Delta J\rangle}{\Delta t}=-\frac{J}{\tau_{\gas}},
\ena
where $\tau_{\gas}$ is the gaseous damping time:
\bea
\tau_{\gas}&=&\frac{3}{4\sqrt{\pi}}\frac{I_{\|}}{n_{\H}m_{\H}
v_{\th}a_{2}^{4}\Gamma_{\|}},\nonumber\\
&=&6.58\times 10^{4} \hat{\rho}\hat{s}^{2/3}a_{-5}\left(\frac{100\K}{T_{\gas}}\right)^{1/2}\left(\frac{30
\cm^{-3}}{n_{\H}}\right)\left(\frac{1}{\Gamma_{\|}}\right) \yr,~~~~\label{eq:taugas}
\ena
where $a_{-5}=a/10^{-5}\cm$, $\hat{s}= s/0.5$, $\hat{\rho}=\rho/3\g\cm^{-3}$. The thermal velocity of a gas atom of mass $m_{\H}$, in a plasma with temperature $T_{\gas}$ and density $n_{\H}$, is $v_{\th}=\left(2k_{\B}T_{\gas}/m_{\H}\right)^{1/2}$.  $\Gamma_{\|}$ is a geometrical parameter, which is equal to unity for spherical grains. This timescale is comparable to the time necessary for the grain to collide with an amount of gas equal to its own mass.

\subsection{Infrared Emission}
Photons emitted by the grain carry away part of the grain's angular momentum, resulting in damping of the grain rotation. The rotational damping rate by infrared emission can be written as
\bea
\tau_{\rm IR}^{-1}=F_{\rm IR} \tau_{\gas}^{-1},
\ena
where $F_{\rm IR}$ is the rotational damping coefficient for a grain of equilibrium temperature $T_{\d}$ (see \citealt{1998ApJ...508..157D}), which is given by
\bea
F_{\rm IR}	=\left(\frac{0.91}{a_{-5}}\right)\left(\frac{u_{\rad}}{u_{\ISRF}}\right)^{2/3}
\left(\frac{30 \cm^{-3}}{n_{\H}}\right)\left(\frac{100 \K}{T_{\gas}}\right)^{1/2}~~~.\label{eq:tauFIR}
\ena
where $u_{\rad}$ is the energy density of the radiation field, and $u_{\rm ISRF}=8.64\times 10^{-13}\erg\cm^{-3}$ is the energy density of the average radiation field in the solar neighborhood as given by \cite{Mezger:1982p4014}.

The total damping rate is then given by
\bea
\tau_{\rm drag}^{-1}=\tau_{\gas}^{-1}+\tau_{\rm IR}^{-1}=\tau_{\gas}^{-1}\left(1+F_{\rm IR}\right).\label{eq:taudamp}
\ena

For large grains (i.e., $a>0.1\mum$), the gaseous damping is dominant, and $\tau_{\rm drag}\approx \tau_{\gas}$. For small grains (i.e., $a\sim 0.01\mum$), the damping by infrared emission becomes dominant for most of the ISM, except for molecular clouds (\citealt{1998ApJ...508..157D}).

Usually, we represent the grain angular momenta and timescales in units of the thermal angular momentum $J_{\th}$ and gaseous damping time $\tau_{\gas}$. The former is given by
\bea
J_{\th}&=&\sqrt{I_{\|}k_{\B}T_{\gas}}=\sqrt{\frac{8\pi\rho s a_{2}^{5}}{15}k_{\B}
T_{\gas}},\nonumber\\
&=&1.05\times 10^{-19}\hat{s}^{-1/3}\hat{\rho}^{1/2}
 a_{-5}^{5/2}\left(\frac{T_{\gas}}{100\K}\right)^{1/2} \g\cm^{2}\s^{-1}.\label{eq:Jth}
\ena
 
\section{Grain alignment by Radiative Torques and H$_{2}$ formation torques}\label{sec21}
Consider a grain subject to an external regular torque $\bGamma$ and a damping torque. The evolution of the grain angular momentum is then governed by the conventional equation of motion:
\bea
\frac{d\bJ}{dt}=\bGamma-\frac{\bJ}{\tau_{\rm drag}},\label{eq:dJdt}
\ena
where $\tau_{\rm drag}$ is the rotational damping time given by Equation (\ref{eq:taudamp}).

The value of the grain's angular momentum in a stationary state, denoted by $J_{\max}$, can be obtained by setting $d\bJ/dt=0$, thus
\bea
J_{\max}=\Gamma_{J}\times \tau_{\rm drag},\label{eq:Jmax}
\ena
where $\Gamma_{J}$ is the the torque component projected onto the direction of $\bJ$.

\subsection{Regular torques arising from H$_2$ formation}\label{sec22}
Among three surface processes proposed by \cite{1979ApJ...231..404P} to drive grains to suprathermal rotation, the formation of H$_{2}$ molecules at catalytic sites was suggested as a dominant mechanism to produce pinwheel torques.

Detailed calculations for the pinwheel torques for a brick-like and a spheroidal grain were presented in \cite{1979ApJ...231..404P}, \cite{1997ApJ...487..248L} and \cite{1997ApJ...484..230L}, respectively. For a brick-like grain with sides $2a_{2}, 2a_{2}$ and height $2a_{1}$ considered in \cite{2009ApJ...695.1457H}, the magnitude of torques due to H$_{2}$ formation that acts to spin up the grain along its symmetry axis takes the following form
\bea
\Gamma_{\H_{2}}&=& r^{2}(r+1)^{1/2}\gamma_{\H}(1-y)n_{\H}\langle v_{\H}\rangle(2a_{1})^{3}
\left(\frac{2m_{\H}E_{\rm kin}}{3\nu}\right)^{1/2},\nonumber\\
&=&\left(\frac{16}{3\pi}\right)^{1/2}r^{2}(r+1)^{1/2}\gamma_{\H}(1-y)n_{\H}(2a_{1})^{3}\left(k_{\B}T_{\gas}\right)^{1/2}
\left(\frac{E_{\rm kin}}{\nu}\right)^{1/2},~~~\label{eq:GammaH2}
\ena
where $r=a_{2}/2a_{1}=1/2s$, $1-y$ with $y=2n(\H_{2})/n_{\H}$ is the fraction of atomic hydrogen, $\gamma_{\H}$ is the conversion efficiency from atomic to molecular hydrogen, and $\langle v_{\H}\rangle=\left(8k_{\B}T_{\gas}/\pi m_{\H}\right)^{1/2}$ is the mean speed of H atoms. Here $\nu=\alpha\times 8r(1+r)(2a_{1})^{2}$ where $\alpha$ is the surface density of active sites, and $E_{\rm kin}$ is the mean kinetic energy of H$_{2}$ molecules escaping from the grain surface (see Appendix \ref{sec:apdxH2} detailed derivation).

For the chosen brick, the effective grain size is given by $4\pi/3 a^{3}=(2a_{1})(2a_{2})^{2}$, and we get $a_{1}=a (\pi s^{2}/6)^{1/3}$. Plugging $a_{1}$ and $\nu$ into Equation (\ref{eq:GammaH2}) we obtain
\bea
\Gamma_{\H_{2}}&=&\left(\frac{2}{3\pi}\right)^{1/2}\left(\frac{\pi}{6}\right)^{2/3}\left(\frac{s^{-1/3}}{8}\right)^{1/2}\nonumber\\
&&\times\gamma_{\H}(1-y)n_{\H}(2a)^{2}\left(k_{B}T_{\gas}\right)^{1/2}\left(\frac{E_{\kin}}{\alpha}\right)^{1/2}.~~~~~~~\label{eq:H2torq}
\ena

Using the typical parameters of the ISM one obtains
\bea
\Gamma_{\H_{2}}= 1.89\times 10^{-29}\hat{s}^{-1/6}\hat{\gamma}_{\H}(1-y)\hat{n}_{\H}\hat{T}_{\gas}^{1/2}a_{-5}^{2}\hat{E}_{\kin}^{1/2}\hat{\alpha}^{-1/2}\g\cm^{2}\s^{-2},\nonumber
\ena
where $\hat{\gamma}_{\H}=\gamma_{\H}/0.2$, $\hat{E}_{\rm kin}=E_{\rm kin}/0.2{\rm eV}$, and $\hat{\alpha}=\alpha/10^{12}\cm^{-2}$.
 
From Equations (\ref{eq:Jmax}) and (\ref{eq:H2torq}), we find the maximum angular momentum of the grain spun up by $\H_{2}$ torques in units of $J_{\th}$ as the following:
\bea
\frac{J_{\max}^{\H_{2}}}{J_{\th}}&\approx&375.4 
(1-y)\hat{\gamma}_{\H}\hat{\rho}^{1/2}\hat{s}^{5/6}\hat{E}_{\rm
kin}^{1/2}\hat{\alpha}^{-1/2}
\nonumber\\
&&\times a_{-5}^{1/2}\hat{T}_{\gas}^{-1/2}
\left(\frac{1}{1+F_{\rm IR}}\right),\label{eq:JmaxH2}
\ena
where $\tau_{\gas}$ and $J_{\th}$ are given by Equations (\ref{eq:taugas}) and (\ref{eq:Jth}).

Equation (\ref{eq:JmaxH2}) shows that H$_{2}$ torques increase with the decreased density of active sites $\alpha$. It means that fewer active sites on the entire grain surface produce stronger H$_{2}$ torques. Although being important for H$_{2}$ torques, the $\alpha$ parameter is not well constrained at the moment.

\cite{1995MNRAS.274..679L} argued that, if there are a few active sites per grain, then the poisoning of active sites by O atoms dominates and is more important than the grain resurfacing, resulting in the suppression of H$_{2}$ torques.  For T$_d>$20~K, he found that the mobility of O atoms leads to rapid poisoning of the active, chemisorption sites and hence produces short lived spin-up torques.  However, more recent laboratory experiments on H$_2$ formation \citep{{1997ApJ...483L.131P}, {1999A&A...344..681P}} indicate that, under cold interstellar conditions, it is dominated by physisorbed particles and is efficient only at $T_{\d}\sim$ 6-10~K for olivine and 13-17~K for amorphous carbon \citep{1999ApJ...522..305K}.  \citet{2004ApJ...604..222C} argued that at higher temperatures H atoms can access chemisorption sites, which would extend the H$_2$ formation to higher temperature (cf. \citealt{2005MNRAS.361..565C}).  Whether the physisorption sites can be localized enough to allow for long-lived "Purcell rockets" is not clear.  If not, it might be that H$_2$ torques can only contribute significantly to grain alignment at relatively high dust temperatures, even under the influence of active site poisoning.  For the present analysis, we will assume that the formulation of, and constraints on, H$_2$ torques from \cite{1995MNRAS.274..679L} is valid. Therefore, small grains with few active sites are unlikely spun up to suprathermal rotation by H$_{2}$ torques. 

In addition, small grains are shown to undergo fast thermal flipping for which systematic torques fixed within the grain body are significantly reduced (see \citealt{1999ApJ...516L..37L}). \cite{2009ApJ...695.1457H} considered the effects of thermal flipping on grain spin-up by pinwheel torques and found that the achievable angular momentum is decreased by a reduction factor $\Delta f$ (see their Figure 13), compared to the maximum value given by Equation (\ref{eq:JmaxH2}). 

Furthermore, the magnitude of $\H_{2}$ torques is determined by the fraction of atomic hydrogen, $1-y=n(\H)/n_{\H}$. The steady fraction of atomic to molecular hydrogen is determined by the equilibrium between the formation and destruction of H$_{2}$ molecules. The H$_{2}$ destruction is dominated by photodissociation due to strong radiation fields from the star. As a result, in dense clouds, very low fraction of atomic hydrogen is expected, while in diffuse clouds, a significant fraction of atomic hydrogen is expected \citep{Rachford:2009hz}. The effect of H$_{2}$ formation on grain alignment is then negligible for the former case.

\subsection{Anisotropic Radiative Torques}\label{sec23}
Let $u_{\lambda}$ be the spectral energy density of radiation field at wavelength $\lambda$ and $\gamma_{\rad}$ its anisotropy. The energy density of radiation field is $u_{\rad}=\int u_{\lambda}d\lambda$. Radiative torque arising from the interaction of an anisotropic radiation field of direction $\bk$ with an irregular grain of size $a$ is then given by
\bea
{\bGamma}_{\lambda}=\gamma_{\rad} \pi a^{2}
u_{\lambda} \left(\frac{\lambda}{2\pi}\right){\bQ}_{\Gamma},\label{eq:GammaRAT}
\ena
where ${\bf Q}_{\Gamma}$ is the RAT efficiency, which can be decomposed into three components $Q_{e1}, Q_{e2}$ and $Q_{e3}$ in a scattering reference frame defined by unit vectors $\ehat_{1},\ehat_{2},\ehat_{3}$ with $\hat{\be}_{1} \| \bk$, $\ehat_{2}\perp \ehat_{1}$ and $\ehat_{3}=\ehat_{1}\times \ehat_{2}$ (\citealt{1996ApJ...470..551D}; \citealt{2007MNRAS.378..910L}).

In general, the magnitude of RAT efficiency $Q_{\Gamma}$ depends on the radiation field, grain shape, size and its orientation relative to $\bk$. When the anisotropic direction of the radiation field is parallel to the axis of maximum moment of inertia $\hat{\ba}_{1}$, LH07 found that the magnitude of RAT efficiency can be approximated by a power-law as:
\bea
Q_{\Gamma}\approx 0.4\left(\frac{{\lambda}}{a}\right)^{\eta},\label{eq:QAMO}
\ena
where $\eta=0$ for $\lambda \ltsim 2a$  and $\eta=-3$ for $\lambda \gg a$. 

From Equations (\ref{eq:Jmax}) and (\ref{eq:QAMO}) one can then determine the maximum angular momentum induced by RATs as:
\bea
\frac{J_{\max}^{\RAT}}{J_{\th}}&=&\left(\int \Gamma_{\lambda} d\lambda\right) \frac{\tau_{\rm drag}}{J_{\th}},\\
&\approx &200\hat{\gamma}_{\rad}\hat{\rho}^{1/2}a_{-5}^{1/2}
\left(\frac{30\cm^{-3}}{n_{\H}}\right)\left(\frac{100\K}{T_{\gas}}\right)\nonumber\\
&&\times
\left(\frac{\bar{\lambda}}
{1.2\mum}\right)\left(\frac{u_{\rad}}{u_{\ISRF}}\right)\left(\frac{\overline{Q_{\Gamma}}}{10^{-2}}\right)
\left(\frac{1}{1+F_{\rm IR}}\right),~~~~~\label{eq:Jmax_RAT}
\ena
where $\hat{\gamma}_{\rad}=\gamma_{\rad}/0.1$, and
\bea
\bar{\lambda}&=&\frac{\int \lambda u_{\lambda} d\lambda}{u_{\rad}},\label{eq:wavemean}\\
\overline{Q}_{\Gamma}&=&\frac{\int Q_{\Gamma} \lambda u_{\lambda}d\lambda}{\overline{\lambda}u_{\rad}},\label{eq:Qmean}
\ena
are the wavelength and RAT efficiency averaged over the entire radiation field spectrum, respectively.

Using Equations (\ref{eq:QAMO})-(\ref{eq:Qmean}), we can calculate $J_{\max}^{\RAT}$ due to RATs for an arbitrary grain of size $a$ embedded in a known radiation field $u_{\lambda}$.

The characteristic timescale for RATs to spin up a grain from thermal to suprathermal rotation is defined as
\bea
\tau_{\rm spin-up}&=&\frac{J_{\th}}{dJ/dt}=\frac{J_{\th}}{\Gamma-J/\tau_{\rm drag}}\nonumber\\
&=&\frac{\tau_{\rm drag}}{J_{\max}/J_{\th}-1},\label{eq:tau_spinup}
\ena
where Equation (\ref{eq:dJdt}) has been used.

The maximum angular momentum that the grain is spun up to, due to both RATs and $\H_{2}$
torques, depends on $a$, $u_{\rad}$, and $\alpha$. Thus, we can write
\bea
J_{\max}(a,\bar{\lambda},u_{\rad},\alpha,y)=J_{\max}^{\RAT}(a,\bar{\lambda},u_{\rad})+J_{\max}^{\H 2}(\alpha, y)\Delta f,~~~~~
\label{eq:Jmaxtot}
\ena
where $\Delta f$ accounts for the reduction of the pinwheel torques due to the grain thermal flipping \citep{2009ApJ...695.1457H}.

\subsection{Dependence of RAT alignment on radiation direction}\label{sec:angleRAT}
The maximum grain angular momentum induced by RATs, $J_{\max}^{\RAT}$ (as given by Equation \ref{eq:Jmax_RAT}), is obtained assuming that $\kv$ is parallel to the axis $\hat{\ba}_{1}$ (i.e., the angle between $\kv$ and $\hat{\ba}_{1}$ is $\Theta=0$). In the presence of an ambient magnetic field, the grain usually rotates about the axis of alignment, $\Bv$. Thus, if $\kv$ is not parallel to $\Bv$, the actual value $J_{\max}^{\RAT}$ is reduced due to projection effects.

\cite{2009ApJ...697.1316H} found that $J_{\max}^{\RAT}$ decreases with an increasing angle $\psi$ between $\kv$ and $\Bv$. Since the RAT alignment tends to occur with $\bJ$ parallel (antiparallel) to $\Bv$, only the RAT component projected onto $\Bv$ spins the grains up to a maximum angular momentum.

Using the analytical expressions of $Q_{e1}$ and $Q_{e2}$ from AMO for the default model, we obtain the following (see Appendix \ref{apdx:Jmax-psi} for derivation) :
\bea
J_{\max}^{\RAT}(\psi) = J_{\max}^{\RAT}(\psi=0)\cos\psi,\label{eq:Jmaxpsi}
\ena
where $J_{\max}^{\RAT}(\psi=0)$ is given by Equation (\ref{eq:Jmax_RAT}).

From Equation (\ref{eq:Jmaxpsi}) we can see that $J_{\max}^{\RAT}(\psi=90^{\circ})=0$. However, this zero value is obtained for the case without internal thermal fluctuations (see Section \ref{sec:Rali}). When such thermal fluctuations are taken into account, it is expected that $J_{\max}^{\RAT}(\psi=90^{\circ}) \sim J_{\rm th}$, i.e., grains rotate thermally regardless of radiation intensity (\citealt{2008MNRAS.388..117H}).

In addition to dependence on $\psi$, the existence of high-$J$ attractor points depend on other parameters, including grain shape, size, and spectrum of the radiation field (see LH07).  Observational evidence supporting a dependence for the grain alignment on the angle $\psi$ was reported in \cite{2010ApJ...720.1045A} and \cite{2011A&A...534A..19A}.

\subsection{Suprathermal rotation and critical size of aligned grains}\label{sec:aali}

In the framework of RAT alignment, some grains are aligned at high-$J$ attractor points with $J_{\hi}=J_{\max}(\psi)$ given by Equation (\ref{eq:Jmax_RAT}), whereas most grains are driven to low-$J$ attractor points having $J_{\lo}\sim J_{\th}$. The alignment of grains at high$-J$ attractor points is stable if grains rotate suprathermally, i.e., $J_{\max}(\psi)\gg J_{\th}$. Using the Langevin equations to follow the RAT alignment of grains in the presence of gaseous randomization, \cite{2008MNRAS.388..117H} found that grains can have nearly stable alignment when $J_{\max}(\psi)/J_{\th}\approx 3$. 

Let $a_{\ali}$ be the critical size of aligned grains, which is taken to be the grain size at which $J_{\max}(\psi)\approx 3J_{\th}$. Since RATs increase rapidly with $a$, grains larger than $a_{\ali}$ would be suprathermally rotating. Using Equation Equation (\ref{eq:Jmax_RAT}) one can determine $a_{\ali}$ as a function of the environment parameters, including  $n_{\H}, T_{\gas}$, and $\bar{\lambda}, u_{\rad}$. 

\subsection{Parameterizing Degree of RAT Alignment}\label{sec:Rali}

Let $Q_{X}=\langle G_{X}\rangle $ with $G_{X}=\left[3\cos^{2}\theta-1\right]/2$ be the degree of internal alignment of the grain axis $\hat{\ba}_{1}$ with $\bJ$, and let $Q_{J}=\langle G_{J}\rangle$ with $G_{J}=\left[3\cos^{2}\beta-1\right]/2$ be the degree of external alignment of $\bJ$ with $\Bv$. Here $\theta$ is the angle between $\hat{\ba}_{1}$ and $\bJ$, and the angle brackets denote the average over the ensemble of grains. The net degree of alignment of $\hat{\ba}_{1}$ with $\Bv$, namely the Rayleigh reduction factor, is defined as $R(a) = \langle G_{X}G_{J}\rangle $. 

Small grains (e.g., $a<a_{\ali}$) can be weakly aligned by paramagnetic relaxation with degree less than $5\%$ for the typical interstellar magnetic fields (\citealt{Hoang:2014cw}). Due to their low mass, small grains mostly produce starlight polarization at ultraviolet wavelengths and have a minor contribution for optical and near-IR starlight polarization. Therefore, we disregard the contribution of the small grains and set $R(a<a_{\ali})=0$. 

While the degree of RAT alignment of the $a\ge a_{\ali}$ grains is not yet available from ab-initio calculations, it can be represented through a set of parameters inferred from the RAT alignment theory. Let $f_{\hi}$ be the fraction of grains that are aligned at high-$J$ attractor points, hence the fraction of grains aligned at low-$J$ attractor points is $1-f_{\hi}$. Since grains at high$-J$ attractor points more likely have perfect alignment of $\bJ$ with $\Bv$, we can write $Q_{J}$ as the following:
\bea
Q_{J}=f_{\hi}+(1-f_{\hi})Q_{J,\lo},\label{eq:QJ}
\ena
where $Q_{J,\lo}$ is the degree of external alignment of grains at low-$J$ attractors.

The Rayleigh reduction factor then becomes:
\bea
R(a)= \langle G_{J}G_{X}\rangle =f_{\hi}+(1-f_{\hi})Q_{J,\lo}Q_{X,\lo},~~~\label{eq:Rayleigh}
\ena
where we use the fact that the external alignment with high-$J$ attractors corresponds
to perfect alignment of $\hat{\ba}_{1}$ with $\bJ$, i.e., $Q_{X,\hi}=1$ due to suprathermal rotation. Equation (\ref{eq:Rayleigh}) can be further simplified using an upper limit $Q_{J,\lo}= 1$. Here, the correlation of $Q_{X}$ and $Q_{J}$ is disregarded, which is minor for suprathermal grains (see \citealt{1999MNRAS.305..615R}). 

The degree of internal alignment, $Q_{X,\lo}$, at the low-$J$ attractor points where the grain axes undergo strong thermal fluctuations due to the vibrational-rotational energy exchange (VRE) can be numerically calculated using adiabatic approximation:
\bea
Q_{X,\lo}=\int_{0}^{\pi} \frac{\left(3\cos^{2}\theta-1\right)}{2}f_{\rm VRE}(\theta, J_{\lo})\sin\theta d \theta,~~~~\label{eq:QXl}
\ena
\bea
f_{\rm VRE}(\theta, J_{\lo})=\mathcal{Z}\exp\left(-\frac{J_{\lo}^{2}}{2I_{\|}k_{\B}T_{\d}}\left[1+(h-1)\sin^{2}\theta\right]\right),~~~\label{eq:fTE}
\ena
is the distribution function of grain axis with respect to the angular momentum. Here $\mathcal{Z}$ is a normalization factor, determined by setting $\int_{0}^{\pi} f_{\rm VRE} \sin\theta d\theta=1$, and $h=I_{\|}/I_{\perp}$ (see \citealt{1997ApJ...484..230L}). 

The exact value of $J_{\lo}$ is uncertain and likely depends on several parameters, including RATs, grain temperature, and gas temperature. Thus, we take a typical value $J_{\lo}= J_{\th}$, which corresponds to the thermal equilibrium between gas randomization and grain rotation. Using Equation (\ref{eq:QXl}) we obtain $Q_{X\lo}(J_{\th}) \approx 0.12$ for oblate grain of $s=0.5$ and $h=1.6$ and the conditions of IC 63. We will see later that $f_{\hi}$ must be $\ge 0.5$ to reproduce observations for IC 63. In this case, reducing $J_{\lo}$ from the typical value $J_{\th}$ only changes the total degree of alignment $R$ by less than $10\%$ because $Q_{X,\lo}$ is much smaller than the first term.

As shown in Equation (\ref{eq:Rayleigh}), the most important parameter in polarization modeling by RATs is the fraction of grains aligned with high-$J$ attractor points, $f_{\hi}$. LH07 showed that $f_{\hi}$ depends on numerous physical parameters, including the grain size $a$, grain shape, $q^{\max}$, anisotropic direction of radiation $\psi$ relative to the magnetic field, and radiation spectrum. For silicate of normal paramagnetism, the RAT alignment tends to have $f_{\hi}<1$. In the case of RAT alignment with high-$J$ attractor points, $f_{\hi}$ may achieve unity (i.e., perfect alignment) when collisional pumping by gas collisions is taken into account. Nevertheless, the RAT alignment with high-$J$ attractors is not universal. It was also found that $f_{\hi}$ can be significantly increased when the ordinary paramagnetic grain has the inclusion of clusters of iron atoms \citep{Lazarian:2008fw}. Therefore, we will not attempt to compute exact values of $f_{\hi}$ but treat it as a model parameter throughout this paper.

\subsection{Theoretical Consideration for Effects of H$_{2}$ Torques}

To better understand the effects of H$_2$ torques on the RAT alignment, let us consider the typical regime of RAT alignment with high-$J$ and low-$J$ attractor points. 

For the alignment at high-$J$ attractor points, if RATs are strong, small grains $a \le 0.05 \mum$ can still rotate suprathermally. Thus, the effect of H$_{2}$ torques for these small grains is minor because the rapid flipping of small grains act to significantly suppress H$_2$ torques (\citealt{1999ApJ...516L..37L}). For weak RATs, only large grains (e.g., $a >a_{\ali}\sim 0.1\mum$) can rotate suprathermally, whereas intermediate grains (i.e., $a \sim 0.05-0.1\mum$) undergo slow flipping (\citealt{1999ApJ...516L..37L}; \citealt{2009ApJ...695.1457H}). In this case, the presence of H$_{2}$ torques helps to drive some intermediate grains to suprathermal rotation, which results in the decrease of $a_{\ali}$ (i.e., smaller grains can still be aligned).

For the alignment at low-$J$ attractor points, it is noted that the spin-up component of RATs is negative in the vicinity of these attractor points, which tends to drive grains to thermal rotation (see LH07). For $a>> a_{\ali}$, the flipping is expected to be slow, and H$_{2}$ torques tend to increase the angular momentum of the low-$J$ attractor points. In this case, a new high-$J$ attractor point will be produced if H$_2$ torques are sufficiently strong to counter the negative spin-up component of RATs (\citealt{2009ApJ...695.1457H}). Therefore, the presence of sufficiently strong H$_2$ torques tends to increase the fraction of grains with high-$J$ attractor points, $f_{\hi}$.

\begin{figure}
\includegraphics[width=0.4\textwidth]{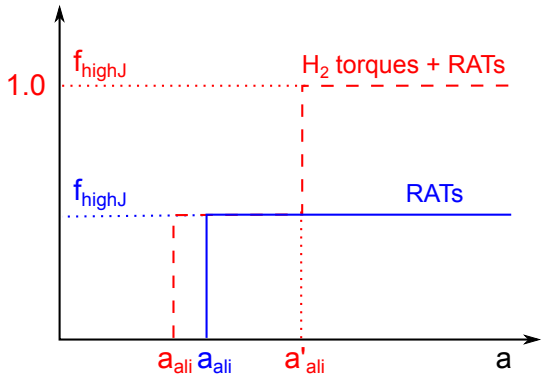}
\caption{Illustration of the variation of the fraction of grains aligned on high-$J$ attractor points $f_{\hi}$ with grain size $a$ for the alignment by RATs (solid line) and by both H$_{2}$ torques and RATs (dashed line). Grains larger than $a_{\ali}$ are aligned with a fraction $f_{\hi}$ on high-$J$ attractor points. The effects of additional H$_2$ torque are to reduce $a_{\ali}$ of grains aligned at high-$J$ and increase $f_{\hi}$ to unity for $a\ge a'_{\ali}$.}
\label{fig:fhighJ}
\end{figure}

In summary, the inclusion of H$_{2}$ torques is expected to result in (i) the decrease of $a_{\ali}$ and (ii) the increase of $f_{\hi}$. The new value $a_{\ali}$ in the presence of H$_2$ torques can be easily calculated as in \S \ref{sec:aali}. To evaluate the increase of $f_{\hi}$ with H$_2$ torques, first we compute the rotation rate of grains at low-$J$ attractor points through Equation (37) in \cite{2009ApJ...695.1457H}, which takes into account the reduction of RATs due to thermal fluctuations and grain flipping. Then, we calculate the minimum size of grains from the low-$J$ attractors that rotate suprathermally, $a'_{\ali}$. Therefore, grains with $a> a'_{\ali}$ should rotate suprathermally, and we assume $f_{\hi}(a> a'_{\ali})=1$. Grains with $a_{\ali}<a<a'_{\ali}$ still rotate thermally at low-$J$ attractor points. Stronger H$_2$ torques tend to result in smaller $a'_{\ali}$, which corresponds to a higher degree of alignment ($f_{\hi}$). The variation of $f_{\hi}$ with $a$ can be summarized in Figure \ref{fig:fhighJ} for the RAT alignment without and with H$_2$ torques.

\section{Quantitative Modeling of RAT alignment and Dust Polarization}\label{sec:IC63}
In this section, we describe the general steps taken to model grain alignment in a molecular cloud induced by RATs plus $\H_{2}$ formation torques, and to predict linear polarization of light from background stars by aligned grains.

\subsection{Physical Model of IC 63 and Model Set up}\label{sec31}

Let us consider an idealized model of the IC 63 nebula, which can be approximated as a spherical cloud of uniform gas density of radius $r_{\rm c}$. The nebula is illuminated by the star $\gamma$ Cassiopeia located at projected distance $d_{\star,\rm sky}=1.3$ pc from the star and the attenuated ISRF. Physical parameters adopted for IC 63, including gas density $n(\H_{2})$, temperature $T_{\gas}$, dust temperature $T_{\d}$ (\citealt{1994A&A...282..605J}; \citealt{1996A&A...309..899J}; \citealt{2005ApJ...628..750F}) and the parameters of $\gamma$ Cas are listed in Table \ref{tab:IC63}.

Based on the geometry of the diffuse emission, \cite{2013ApJ...775...84A} found that IC 63 and $\gamma$ Cas are offset from each other along the sightline, with the line between $\gamma$ Cas and IC 63 making an angle of $\gamma_{\star}=58^{\circ}$ with respect to the plane of the sky (henceforth POS). The actual distance from $\gamma$ Cas to IC 63 is $d_{\star}=d_{\star,\rm sky}/\cos(58^{\circ})=2.45$pc. In addition, both the position angles of the optical polarization in the nebula and synchrotron measurements in the area encompassing IC~63 \citep{2007A&A...463..993S}  show a projected magnetic field closely parallel to the Galactic plane. While direct determinations of the three dimensional structure of the magnetic field in the region are not available, we can estimate the over-all field orientation by assuming that it follows the spiral arms.  If we use a pitch angle for the local and Persus arms of $\sim10^\circ$ \citep{2013ApJ...769...15X}, and the Galactic longitude for IC~63 of \textit{l}=163$^\circ$, we find that the magnetic field makes an angle of $\sim 43^\circ$ with the sightline (or an angle $\xi=47^{\circ}$ with the POS).

\begin{figure}
\includegraphics[width=0.4\textwidth]{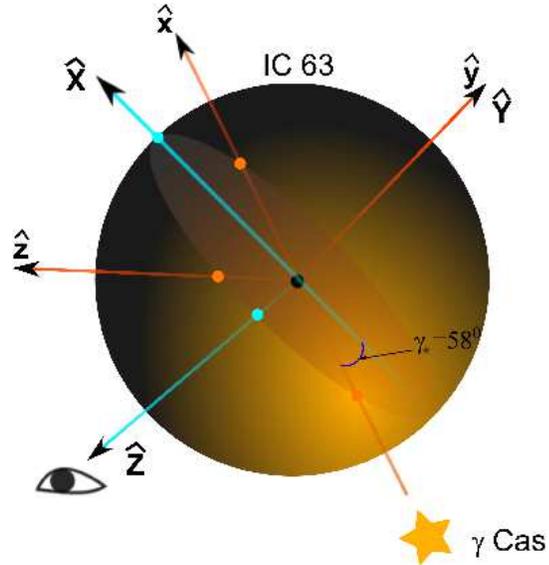}
\caption{3D schematic illustration of the IC 63 nebula and the $\gamma$ Cas star, and coordinate systems defined for calculations. $\Zhat$ denotes the sightline toward background stars, and the $\Xhat\Yhat$ plane perpendicular to $\Zhat$ describes the POS. $\gamma$ Cas illuminates IC 63 in the $\xhat$ direction, which is along the path connecting the star and nebula, and the star-nebula-path makes an angle $\gamma_{\star}=58^{\circ}$ with the POS $\Xhat\Yhat$. The $\xhat\zhat$ plane with $\zhat \perp \xhat$ describes a cut through the nebula containing the star-nebula-path and perpendicular to the POS. The cyan and orange filled circles denote the intersections of the vectors with the nebula.}
\label{fig:IC63}
\end{figure}

For our modeling, we define two coordinate systems centered in IC 63, $\xhat\yhat\zhat$ and $\Xhat\Yhat\Zhat$ with $\yhat \equiv \Yhat$, where the $\Xhat\Yhat\Zhat$ coordinate system is obtained by rotating the $\xhat\yhat\zhat$ system by an angle $\gamma_{\star}$ around the  $\Yhat$ axis (see Figure \ref{fig:IC63}).

\begin{table}
\centering
\caption{Physical parameters for the IC 63 nebula}\label{tab:IC63}
\begin{tabular}{l c }\hline\hline\\
Parameters & Values \cr
\hline
Nebula radius, $r_{\rm c}$ & $0.03\pc$\cr
n($\H_{2}$)& {$2-4\times 10^{4}\cm^{-3}$}\cr
{$n(\H)/2n(\H_{2})$ near surface} & {0.1}  $^{\rm a}$\cr
{$T_{\gas}$}& {$150\K$}\cr
{$T_{\d}$}& {$45\K$}\cr
{$\gamma$ Cas}& {\B0.5 \rm IV}\cr
{Nebula-$\gamma$ Cas projected distance, $d_{\star,\rm sky}$} &{$1.3\pc$}\cr  
{$R_{\star}$} &{$14R_{\odot}$}\cr  
{$T_{\star}$} &{$30500\K$}\cr
{Distance of the star, D} &{200 pc}\cr
{H$_2$ formation efficiency, $\gamma_{\H}$} & {$0.1$}\cr 
\cr
\hline
\cr
\multicolumn{2}{l}{$^{\rm a}$ see
\cite{2005ApJ...628..750F}.}
\end{tabular}
\end{table}

\subsection{Simplified Radiative transfer}\label{sec32}
Dust grains inside IC 63 are illuminated both by stellar radiation from $\gamma$ Cas as well as the attenuated interstellar radiation field (ISRF). For the latter, we adopt the radiation field from \cite*{1983A&A...128..212M} (hereafter MMP), where the mean radiation intensity $J_{\lambda}^{\rm MMP}$ inside a giant molecular cloud located at 5 kpc from the Galactic center is calculated for different visual extinctions $A_{V}^{\rm MMP}$ from the surface (see also \citealt{2005ApJ...631..361C}). A degree of anisotropy $\gamma_{\rad}=0.35$ is assumed, as numerically calculated in \cite{2007ApJ...663.1055B}. The stellar radiation from $\gamma$ Cas is completely anisotropic with $\gamma_{\rad}=1$. Since the distance between the star and nebula is much larger than the nebula's radius, the incident stellar radiation can be approximated as parallel beams.

\begin{figure}
\includegraphics[width=0.4\textwidth]{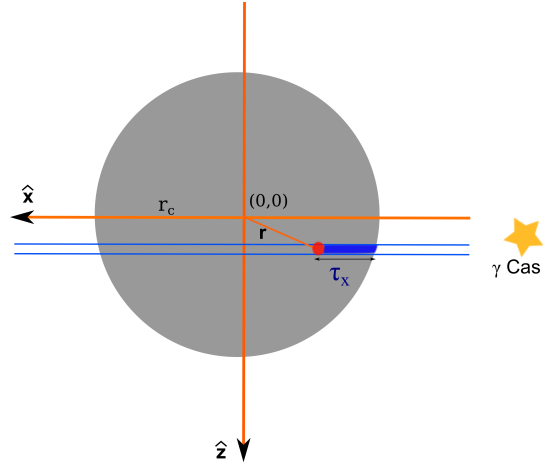}
\caption{Schematic of grain alignment in a circular slab $\xhat\zhat$ with y=0 (middle slab) by the stellar radiation from $\gamma$ Cas. $r$ is the distance from the grain to the center of IC 63 and $\tau_{x}$ is the optical depth with respect to $\gamma$ Cas. The positive direction of $x$ is chosen along the direction of light propagation. Dayside and nightside have $x<0$ and $x>0$, respectively.}
\label{fig:IC63-2D}
\end{figure}

For convenience, we begin with a middle slab $\xhat\zhat$ of $y=0$ as shown in Figure \ref{fig:IC63-2D}, which contains both the IC 63-$\gamma$ Cas connecting line and perpendicular to the POS. Let $\tau_{x}$ be the optical depth for the extinction of the stellar radiation by dust along the $\xhat$ direction. We divide the $\xhat\zhat$ plane into $N_{x}\times N_{z}$ cells with $N_{x}=N_{z}=128$. The gas density, temperature, and density of radiation energy at each cell are given by $n_{\H}(x,z)$, $T_{\gas}(x,z)$, and $u_{\lambda}(x,z)$, respectively.

Provided the effective temperature $T_{\star}$ of $\gamma$ Cas and the distance $d_{\star}$ from the star to IC 63, we can derive the spectral energy density $u_{\lambda}$ as follows:
\bea
u_{\lambda}=\frac{L_{\lambda}}{4\pi d_{\star}^{2}c}=\frac{4\pi R_{\star}^{2}F_{\lambda}(T_{\star})}{4\pi d_{\star}^{2}c}=\frac{\pi B_{\lambda}(T_{\star})}{c}\left(\frac{R_{\star}}{d_{\star}}\right)^{2}
\ena
where $L_{\lambda}$ is the spectral luminosity, and $F_{\lambda}=\pi B_{\lambda}$ is the spectral flux at the surface of the star (assuming the star is a black body, see e.g., \citealt{1986rpa..book.....R}). Due to the extinction by dust and accounting for the attenuated ISRF, the energy density inside the cloud is given by
\bea
u_{\lambda}=\frac{\pi B_{\lambda}(T_{\star})}{c}\left(\frac{R_{\star}}{d_{\star}}\right)^{2}e^{-\tau_{x}(\lambda)}
+u_{\lambda}^{\rm MMP}(A_{V}^{\rm sd}),~~~\label{eq:urad}
\ena
where the reemission of dust with temperature $T_{\rm d}= 45\K$ (see Table \ref{tab:IC63}) is ignored because the RAT efficiency $Q_{\Gamma}$ decreases subtantially at infrared wavelengths $\lambda>>a$. The second term denotes the attenuated ISRF, which is subdominant for the conditions of IC 63 (see Eq. \ref{eq:Jmax_RAT}). Here $u_{\lambda}^{\rm MMP}(A_{V}^{\rm sd}) = 4\pi J_{\lambda}^{\rm MMP}(A_{V}^{\rm sd})/c$ denotes the energy density at distance determined by visual extinction $A_{V}^{\rm sd}$ from the cloud surface to the dust grain's location, which is obtained by interpolating $J_{\lambda}^{\rm MMP}$ as a function of $A_{V}^{\rm MMP}$ for $A_{V}^{\rm sd}$.

The optical depth by a dust column of thickness $dx$ along the $\xhat$ direction is given by
\bea
d\tau_{x}(\lambda)=\int_{a_{\min}}^{a_{\max}}C_{\ext}(\lambda,a)\left(\frac{dn}{da}da\right)dx,\label{eq:dtaux}
\ena 
where $C_{\ext}=\left(2C_{\perp}+C_{\|}\right)/3$ is the extinction cross-section due to randomly oriented grains (see Appendix \ref{apdx:Cext}), and $dn/da$ is the grain size distribution. Here, the integration over the grain size is carried out for both silicate and carbonaceous grains. By convention, $C_{\|}$ and $C_{\perp}$ are the extinction cross-section for the electric field of incident radiation parallel and perpendicular to the grain symmetry axis, respectively. 

Using the physical parameters for IC 63 in Table \ref{tab:IC63} for Equations (\ref{eq:dtaux}), one can calculate the optical depth as a function of wavelength. With $\tau_{x}$ known, one can calculate $u_{\lambda}$ inside the nebula using Equation (\ref{eq:urad}). The energy density ($u_{\rad}(x,z)$) of the radiation inside the cloud is obtained by integrating over the entire spectrum of $u_{\lambda}$. The mean wavelength ($\bar{\lambda}(x,z)$) is evaluated using Equation (\ref{eq:wavemean}).

\subsection{Critical size of aligned grains}

Using $u_{\rad}(x,z)$, $\bar{\lambda}(x,z)$ together with $n_{\H}$ and $T_{\gas}$ for Equations (\ref{eq:Jmaxtot})-(\ref{eq:Jmaxpsi}), one obtains $J_{\max}/J_{\th}$ due to RATs and H$_2$ torques. The critical size of aligned grains, $a_{\ali}(x,z)$, can be evaluated using the criteria in Section \ref{sec21}. 

\subsection{Linear Polarization}\label{sec33}

Consider a column of dust along the $\Zhat$ axis toward a background star with the visual extinction $A_{V}=1.086\tau_{Z}(\lambda=0.55\mum)$, where $\tau_{Z}$ is obtained by integrating Equation (\ref{eq:dtaux}) along the $\Zhat$ axis. It is noted that the sightlines toward background stars, parallel to the $\Zhat$ axis, make an angle $\gamma_{\star}$ with the $\zhat$ axis (see Figure \ref{fig:IC63}). 

The degree of linear polarization arising from the aligned asymmetric silicate grains in a cell of thickness $dZ$ is computed as the following:
\bea
dp_{\lambda}(X,Z)=\int_{a_{\ali}(X,Z)}^{a_{\max}}\frac{\left(C_{X}-C_{Y}\right)}{2}\frac{dn}{da}dadZ,
\label{eq:dplam}
\ena
where $a_{\ali}(X,Z)$ is the critical size of aligned grains in cell $(X,Z)$ (see Appendix \ref{sec:apdxa} for derivation).

The value $a_{\ali}(X,Z)$ is interpolated from $a_{\ali}(x,z)$ with the use of the coordinate transformations from $\xhat\zhat$ to $\Xhat\Zhat$ coordinate systems. 

As shown previously \citep{{2013ApJ...779..152H}, {Hoang:2014cw}}, small grains have very low, but finite residual alignment degree. However, they only affect starlight polarization in the ultraviolet wavelengths, whereas the peak of polarization spectrum is mostly determined by aligned large grains. Thus, the minor contribution of $a<a_{\ali}$ grains is disregarded in the above equation. 

Equation (\ref{eq:dplam}) can be rewritten as
\bea
dp_{\lambda}(X,Z)=\int_{a_{\ali}(X,Z)}^{a_{\max}} \frac{C_{\pol}}{2}R(a)\cos^{2}\xi
\frac{dn}{da}dadZ,\label{eq:dpRAT}
\ena
where $C_{\pol}=\left(C_{\|}-C_{\perp}\right)$ is the polarization cross-section for oblate spheroidal grains, $\xi$ is the angle between the magnetic field and the POS, and $R(a)$ is given by Equation (\ref{eq:Rayleigh}).

Equations (\ref{eq:dpRAT}) is integrated over $Z$ to obtain the polarization $p_{\lambda}(X)$ for each sightline (i.e., $A_{V}$) with the use of the step function $f_{\hi}$ from Figure \ref{fig:fhighJ}. The transition of alignment from unaligned, small grains to aligned grains at $a=a_{\ali}$ is unlikely a sudden jump (see e.g, \citealt{Hoang:2014cw}), and we multiply $R(a)$ by a smoothing function $f_{\rm sm}=1-\exp\left[-\left(a/a_{\ali} \right)^{3}\right]$.

The extinction cross-section $C_{\ext}$ and polarization cross-section $C_{\pol}$ are taken from \cite{2013ApJ...779..152H} who computed the cross-sections for silicate and carbonaceous grains with the dielectric functions from \cite{2003ApJ...598.1026D}. Oblate spheroidal grains of axial ratio $r=2$ are considered.

\section{Results}\label{sec:results}
In this section, we present predictions for grain alignment and linear polarization, assuming a mixture model of dust consisting of amorphous silicate grains and carbonaceous grains. The grain size distributions are taken from the model with a typical total-to-selective extinction $R_{V}=3.1$ for the ISM from \cite{2001ApJ...548..296W}, in which the distribution for silicate has a sharp decline at $a_{\max}\sim 0.25\mum$. Photometric analysis in \cite{2013ApJ...775...84A} shows that $R_{V}$ varies across IC 63 with an averaged value $\langle R_{V}\rangle =2.27$, which is smaller than the adopted $R_{V}$. For the calculations in this paper, the carbonaceous grains are assumed to be unaligned and thus to not contribute to the polarization \citep{2006ApJ...651..268C}. We consider four different cloud models with gas density $n(\H_{2})=5\times 10^{4}\cm^{-3}$ (model 1), $4\times 10^{4}\cm^{-3}$ (model 2), $3\times 10^{4}\cm^{-3}$ (model 3) and $2\times 10^{4}\cm^{-3}$ (model 4). Results presented here are calculated for $f_{\hi}^{\RAT}=0.5$, i.e., $50\%$ of grains are radiatively aligned with high-$J$ attractor points.  We first model the grain alignment solely by RATs, and subsequently consider the effects of the addition of H$_2$ formation torques.

\subsection{Grain Alignment by RATs}
 \subsubsection{Two-dimensional case}
{\it (a) Isothermal cloud model}

Let us consider an isothermal cloud of gas temperature $T_{\gas}=150\K$ (see Table \ref{tab:IC63}). We present results for grain alignment in a circular slab $\xhat\zhat$ ($\Xhat\Zhat$) at $y=0$ (middle slab through the IC 63 center) considered in the previous section.

\begin{figure}
\includegraphics[width=0.4\textwidth]{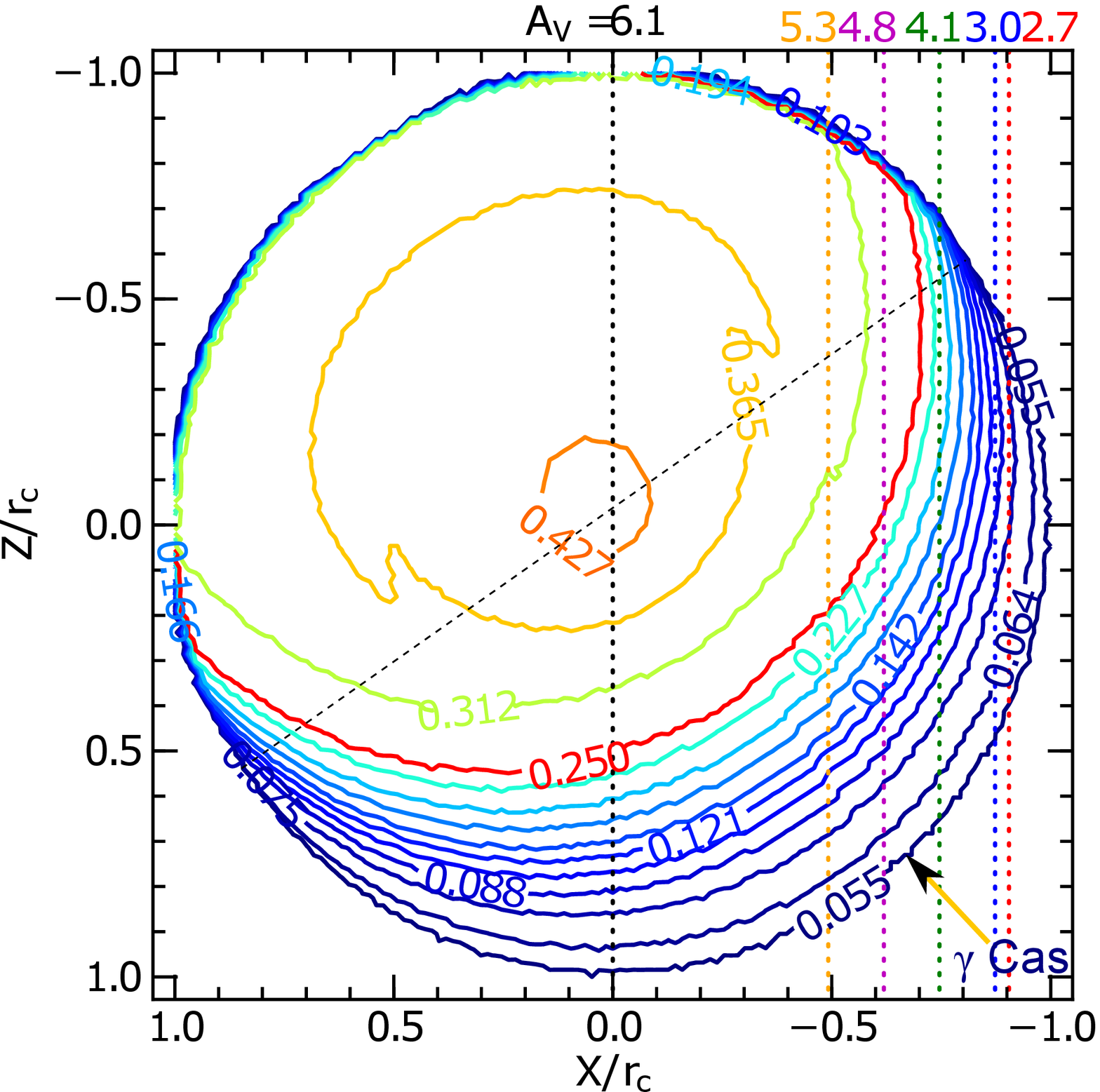}
\includegraphics[width=0.4\textwidth]{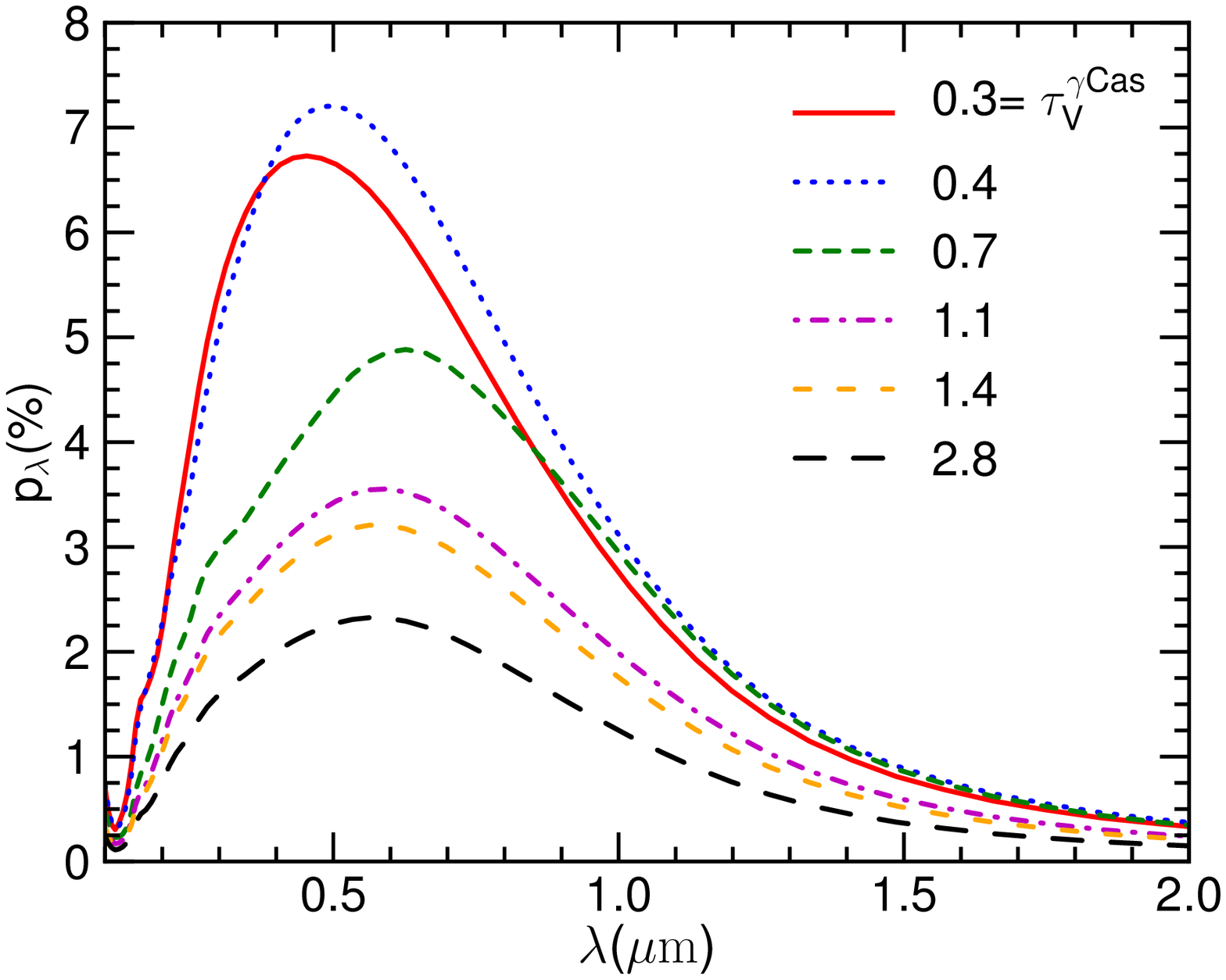}
\caption{{\it Upper panel:} contours of $a_{\ali}$ in microns in the $\Xhat\Zhat$ plane for model 3 with $n(\H_{2})=3\times 10^{4}\cm^{-3}$, where the red contour corresponds to $a_{\ali}\sim a_{\max}$. Stellar radiation arrives from $\gamma$ Cas in the $\xhat$ direction (marked by arrow) that makes an angle $\gamma_{\star}=58^{\circ}$ with $\Xhat$. Dashed diagonal line divides the dayside and nightside. Grains are efficiently aligned in the outer layer of the dayside that face $\gamma$ Cas, and only very large grains can be aligned in the nightside. Dotted lines show several sightlines toward background stars with $A_{V}$ indicated. {\it Lower panel:} polarization curves due to aligned grains by RATs along the selected sightlines shown in the upper panel. The V-band optical depths relative to $\gamma$ Cas in the $\Xhat$ direction, $\tau_{V}^{\gamma \rm Cas}$, are indicated for the selected sightlines. Grain alignment by only RATs for model 3 is considered.}
\label{fig:aali}
\end{figure}

Figure \ref{fig:aali} (upper panel) shows contours of the critical size of aligned grains $a_{\ali}$ induced by RATs only. As shown, in the region facing toward $\gamma$ Cas (hereafter dayside), which are directly illuminated by stellar radiation, grains as small as $a= 0.055\mum$ can be aligned. In the region facing away from $\gamma$ Cas (hereafter nightside), only big grains can be aligned. The critical size $a_{\ali}$ appears to exceed the upper cut-off of grain size distribution ($a_{\max}\sim 0.25\mum$), hence, no aligned grains are present in the nightside, except very thin outer layer where the attenuated ISRF has some effect. That is a direct consequence of the weakening of RATs due to the reddening of stellar radiation while propagating deeper into the nebula.

It is useful to define a column of aligned grains, which is a column of dust that contains all aligned grains with size $a_{\ali}\le a_{\max}$. From Figure \ref{fig:aali} (upper panel), one can see that, for the sightlines passing the region of aligned grains with $X< -0.7$, the column density of aligned grains increases with increasing total dust column density. Beyond $X\sim -0.7$, the column density of aligned grains decreases with increasing the total column density. Here the sightline $X\sim -0.7$ is nearly tangential to the contour $a_{\ali}= 0.25\mum$ (red contour).

Figure \ref{fig:aali} (lower panel) shows the polarization curves calculated for the several sightlines through the nebula with $X \approx -0.9, -0.87,-0.75, -0.6, -0.5$ and $X=0$. For model 3, the visual extinctions of these sightlines are $A_{V}=2.7, 3.0, 4.1, 4.8, 5.3$, and $6.1$. The corresponding V-band optical depths relative to $\gamma$ Cas along the $\Xhat$ axis are $\tau_{V}^{\gamma \rm Cas}=0.3, 0.4, 0.7, 1.1, 1.5$, and $2.8$ (see Eq. \ref{eq:AV_to_tauV}). First, the peak polarization increases with the increasing optical depth up to $\tau_{V}^{\gamma \rm Cas}=0.4$ (i.e., $A_{V}=3.0$), and then it substantially decreases for $\tau_{V}^{\gamma \rm Cas}>0.4$. This can be easily understood. Indeed, Figure \ref{fig:aali} (upper panel) shows that the first three sightlines go through the region with most grains aligned (i.e., $a_{\ali}< 0.1$) while the last three sightlines go through the region with only large grains aligned (i.e., $a_{\ali}> 0.1 $ for large $Z$ only) and lower amount of aligned grains. The wavelength at which the polarization peaks, $\lambda_{\max}$, increases with the increasing $\tau_{V}^{\gamma \rm Cas}$.


{\it (b) Effects of gas temperature variation}

To evaluate the effects of gas temperature variation within the nebula and the resultant variable collision rate of grain alignment, we reran our models with the temperature profile obtained from a two-dimensional model in \cite{1995A&A...302..223J}.  Specifically, they showed a decrease in the gas temperature, $T_{\gas}$, with increasing optical depth in the direction of the stellar illumination, denoted by $\tau_{V,x}^{\gamma \rm Cas}$ (i.e., going deeper into the nebula). For instance, the temperature as high as $T_{\gas}\sim 250\K$ is found at $\tau_{V,x}^{\gamma \rm Cas}\sim 0.2$ (i.e., in the cloud surface facing $\gamma$~Cas) and falls to $T_{\gas} < 150\K$ at $\tau_{V,x}^{\gamma \rm Cas} > 1$ and $T_{\gas} \sim 45\K$ at $\tau_{V,x}^{\gamma \rm Cas} \sim 3.0$. A high gas temperature close to the cloud surface facing $\gamma$ Cas is directly supported by the H$_2$ excitation measurements of \citet{2010ApJ...725..159F}.

\begin{figure}
\includegraphics[width=0.4\textwidth]{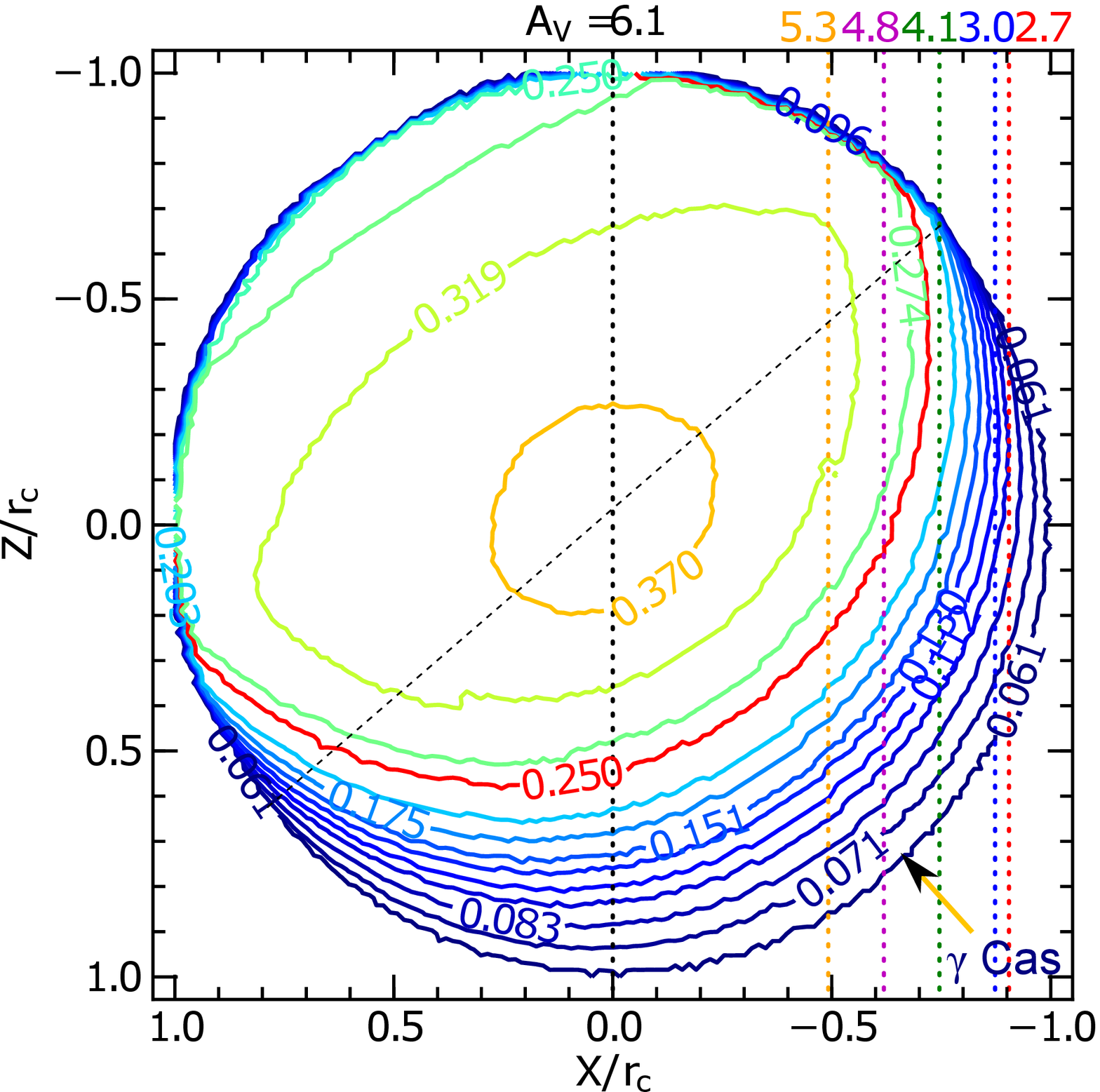}
\includegraphics[width=0.4\textwidth]{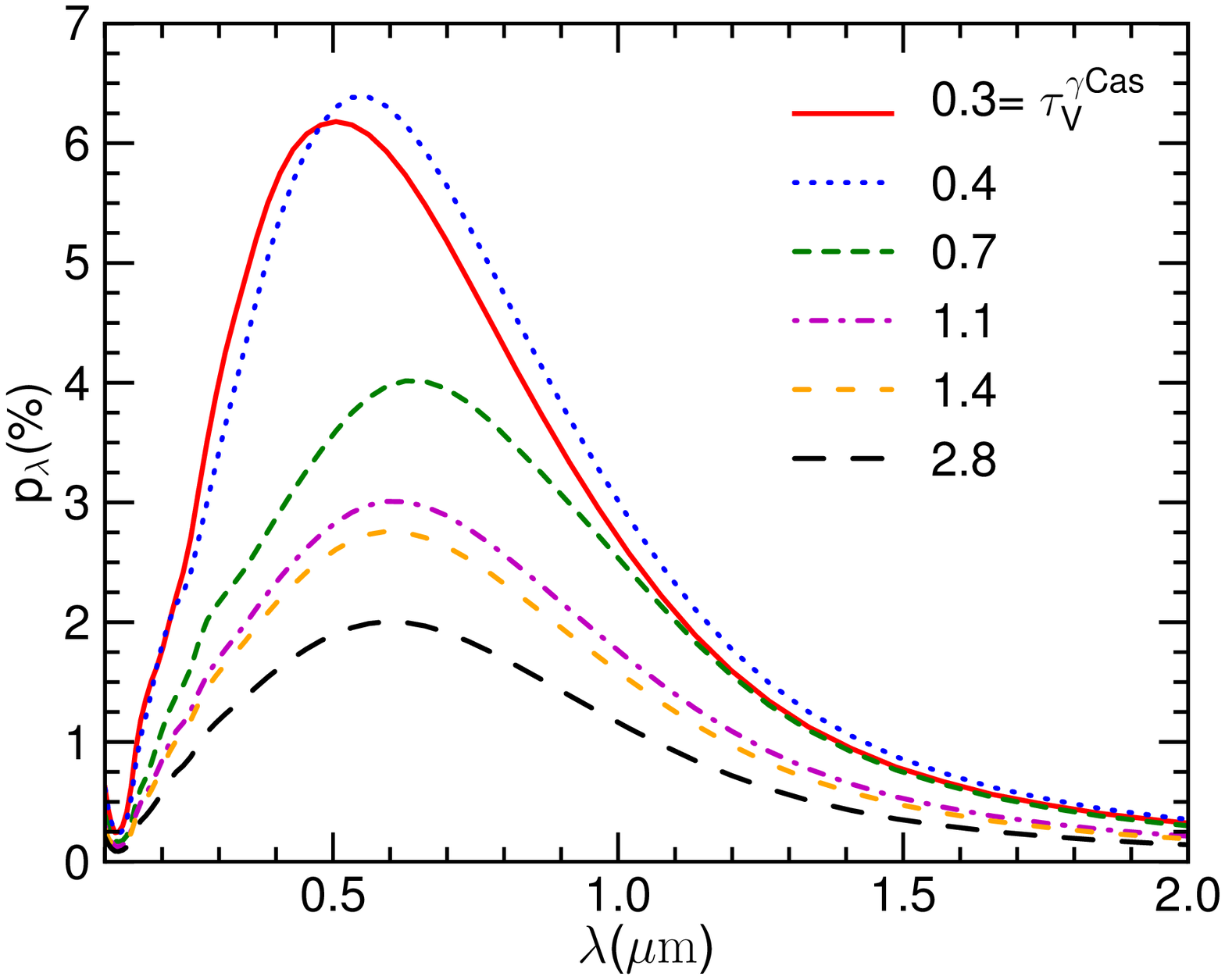}
\caption{Similar to Figure \ref{fig:aali}, but for the case when the gas temperature $T_{\gas}$ decreases with increasing the depth into the cloud.}
\label{fig:aali_Tgas}
\end{figure}

Figure \ref{fig:aali_Tgas} (upper panel) shows the contours of $a_{\ali}$ in the presence of gas temperature variation. Compared to the results for isothermal case in Figure \ref{fig:aali} (upper panel), $a_{\ali}$ increases in the outer layer but decreases in the inner region.
 
The same as the upper panel but Figure \ref{fig:aali_Tgas} (lower panel) shows the polarization curves. For all selected sightlines, the maximum polarization $p_{\max}$ is decreased by a factor of $1.2$ while the wavelength $\lambda_{\max}$ is slightly increased compared to the results for the isothermal model. These results are easy to understand. Indeed, from the upper panel we can see that, along the considered sightlines, most aligned grains are located in the outer regions with $\tau_{V,x}^{\gamma \rm Cas}<1$ ($T_{\gas}> 150\K$), which corresponds to stronger disalignment and produces lower polarization.

{\it (c) Variation of $p/A_{V}$ from the IC 63 edge to center}

\begin{figure}
\includegraphics[width=0.4\textwidth]{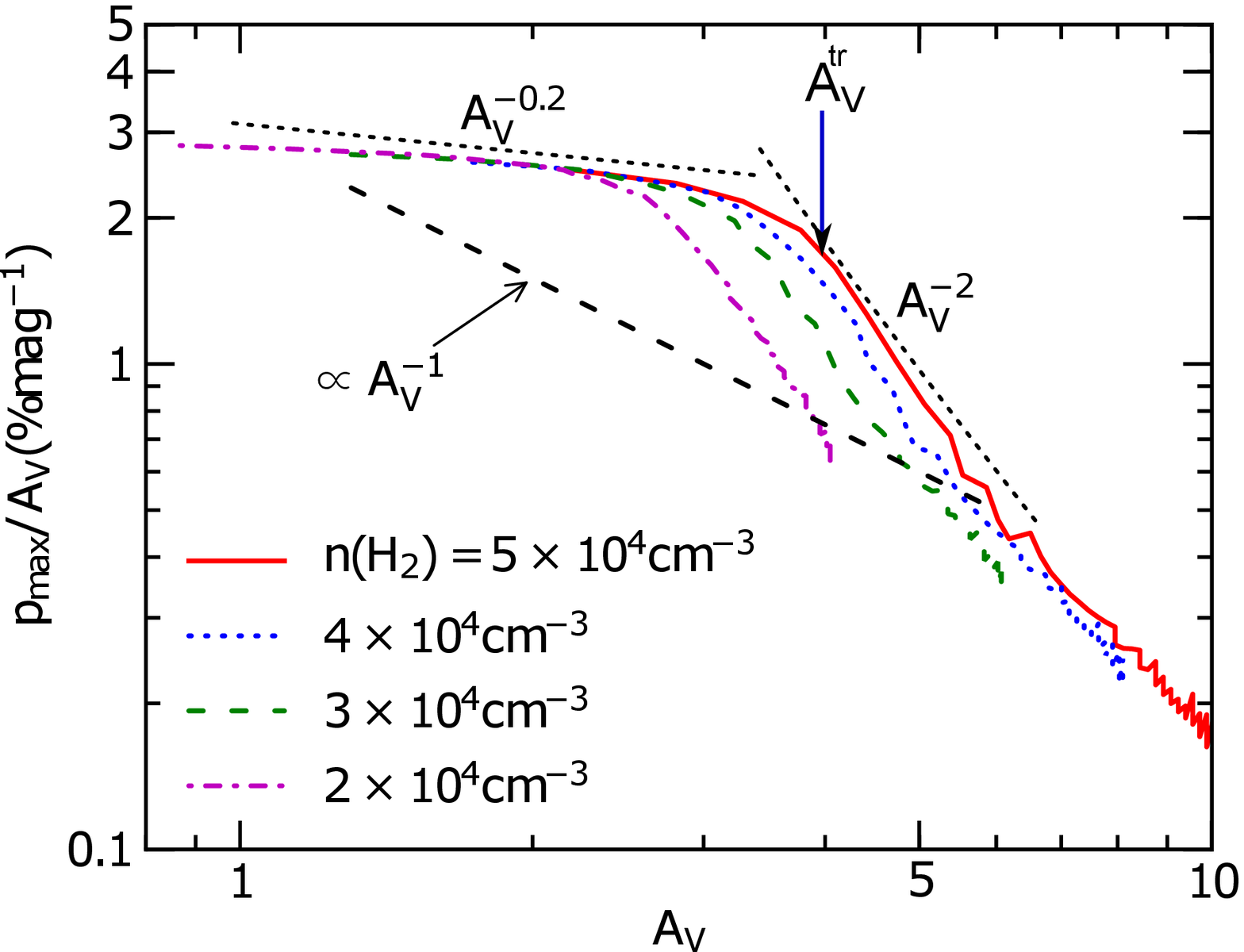}
\includegraphics[width=0.4\textwidth]{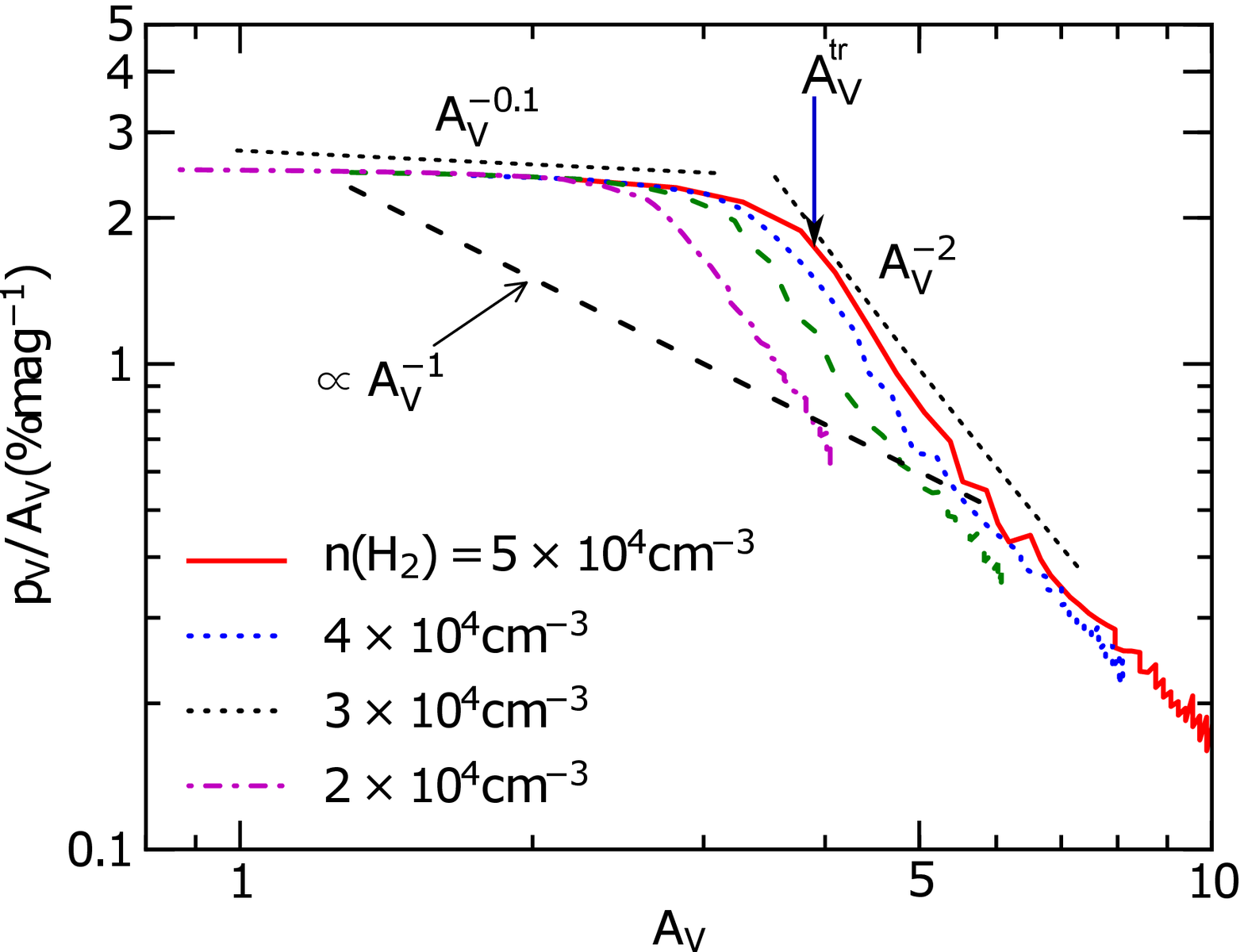}
\caption{Variation of fractional polarization $p/A_{V}$ as a function of $A_{V}$ for four models of IC 63 (see the text). The upper and lower panels show $p_{\max}/A_{V}$ and $p_{V}/A_{V}$, respectively. Two-stage variation of $p/A_{V}$ vs. $A_{V}$ is observed with a shallow slope for $A_{V}<3$ and steep slope for $A_{V}>4$. Dotted lines show the approximate power laws of $p/A_{V}$ vs. $A_{V}$, and arrows mark the transition position with $A_{V}^{tr}$. The dashed line shows $p/A_{V}\propto A_{V}^{-1}$. Grain alignment by only RATs is considered.}
\label{fig:pmaxAV}
\end{figure}

Figure \ref{fig:pmaxAV} shows the variation of fractional polarizations, $p_{\max}/A_{V}$ (upper) and $p_{V}/A_{V}$ (lower), with $A_{V}$, calculated for the four models of IC 63. The variation of $p_{\max}/A_{V}$ with $A_{V}$ essentially follows two separate stages, which can be well described by a power-law $A_{V}^{\eta}$. In the first stage, $p_{\max}/A_{V}$ decreases slowly with $A_{V}$ with a shallow slope $\eta \sim -0.2$. In the second stage, it declines more rapidly, with a very steep slope $\eta \sim - 2$. The variation of $p_{V}/A_{V}$ versus $A_{V}$ is similar to that of $p_{\max}/A_{V}$ in the second stage, but it is slightly shallower in the first stage, with $p_{V}/A_{V}\propto A_{V}^{-0.1}$.

The transition from the shallow slope to steep slope occurs at visual extinction $A_{V}^{\rm tr}$. This transition value tends to decrease with decreasing the gas density. For instance, $A_{V}^{\rm tr}\sim 4$ for $n(\H_{2})=5\times 10^{4}\cm^{-3}$ and decreases to $A_{V}^{\rm tr}\sim 2.5$ for $n(\H_{2})=2\times 10^{4}\cm^{-3}$ (see Figure \ref{fig:pmaxAV}). Interestingly, the very steep slope begins from $A_{V}^{\rm tr}$ where grain alignment has not yet decreased significantly, i.e., the lines of sight $A_{V}\sim A_{V}^{\rm tr}$ still go through the region with $a_{\ali}\ll a_{\max}$ (see Figure \ref{fig:aali_Tgas} for model 3).

\subsubsection{Three-dimensional case}
Due to its spherical symmetry, the three-dimensional cloud is equivalent to the superimposition of $N_{y}$ circular slabs with the slab radius decreasing with increasing $|y|=0$. With the assumption of parallel radiation beams from $\gamma$ Cas and uniform gas density, calculations for $a_{\ali}$ and linear polarization $p_{\lambda}$ for the three-dimensional cloud can be obtained using the results from the two-dimensional case.

To this end, in the two-dimensional case, we have also created data tables of $a_{\ali}(x,y=0,z)$ and $dp_{\lambda}(x,y=0,z)$ calculated by Equation (\ref{eq:dpRAT}) using $a_{\ali}(x,z)$. Then, we interpolate for $dp_{\lambda}(x,y,z)$ for a cell at $y$ coordinate using the dependence of $dp_{\lambda}(x,y=0,z)$ on $\tau_{x}(x,y=0,z)$ with regard to $\gamma$ Cas. For example, for a given coordinate $z$ in the cloud, we have a $xy$ grid. Since $\tau_{x}(x,y=0,z)$ is already calculated, one can easily calculate $\tau_{x}(x,y,z)$ for each cell $(x,y)$ of thickness $dz$. Then, $dp_{\lambda}(x,y,z)$ is obtained by interpolating $dp_{\lambda}(x,y=0,z)$ as a function of $\tau_{x}(x,y=0,z)$ for a given $\tau_{x}(x,y,z)$. Finally, we interpolate for $dp_{\lambda}(X,Y,Z)$ using $dp_{\lambda}(x,y,z)$ and the coordinate transformation from $\xhat\yhat\zhat$ to $\Xhat\Yhat\Zhat$. Polarization in the POS, $p_{\lambda}(X,Y)$, is obtained by integrating $dp_{\lambda}(X,Y,Z)$ along the $\Zhat$ axis.
 
\begin{figure}
\includegraphics[width=0.4\textwidth]{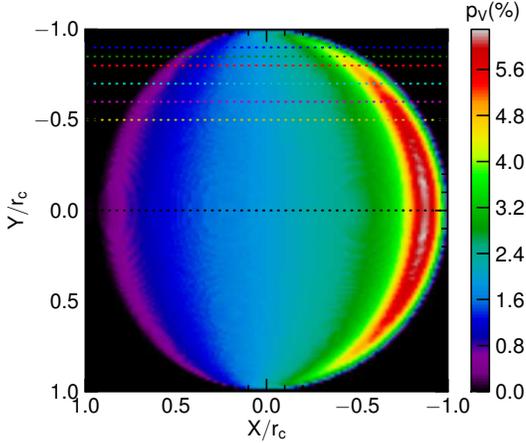}
\caption{Map of $p_{V}$ in the POS, $\Xhat\Yhat$. Stellar radiation projected onto the POS is parallel to the $\Xhat$ axis (from negative to positive X). Dotted lines show the different cuts ($Y/r_{c}=0.0, -0.5, -0.6,-0.7, -0.8, -0.85$ and $-0.9$) along $\Xhat$.The polarization in the dayside is stronger than in the nightside due to its stronger radiation intensity. Along a horizontal cut (constant $Y$), $p_{V}$ first increases to its maximum and then declines substantially toward the night side.}
\label{fig:pVcontour}
\end{figure}

Figure \ref{fig:pVcontour} shows the map of $p_{V}(X,Y)$ in the POS. As expected from the maps of grain alignment (Figure \ref{fig:aali}), the polarization is stronger in the dayside of the cloud and decreases significantly toward the nightside. The polarization appears to peak in the region very close to the cloud surface (red), extending from $Y/r_{c}=-0.7$ to $+0.7$. On the nightside of IC 63, the stellar radiation is significantly reduced due to dust extinction and the fraction of aligned grains decreases, which results in the substantial decrease of polarization. 

\begin{figure*}
\includegraphics[width=0.45\textwidth]{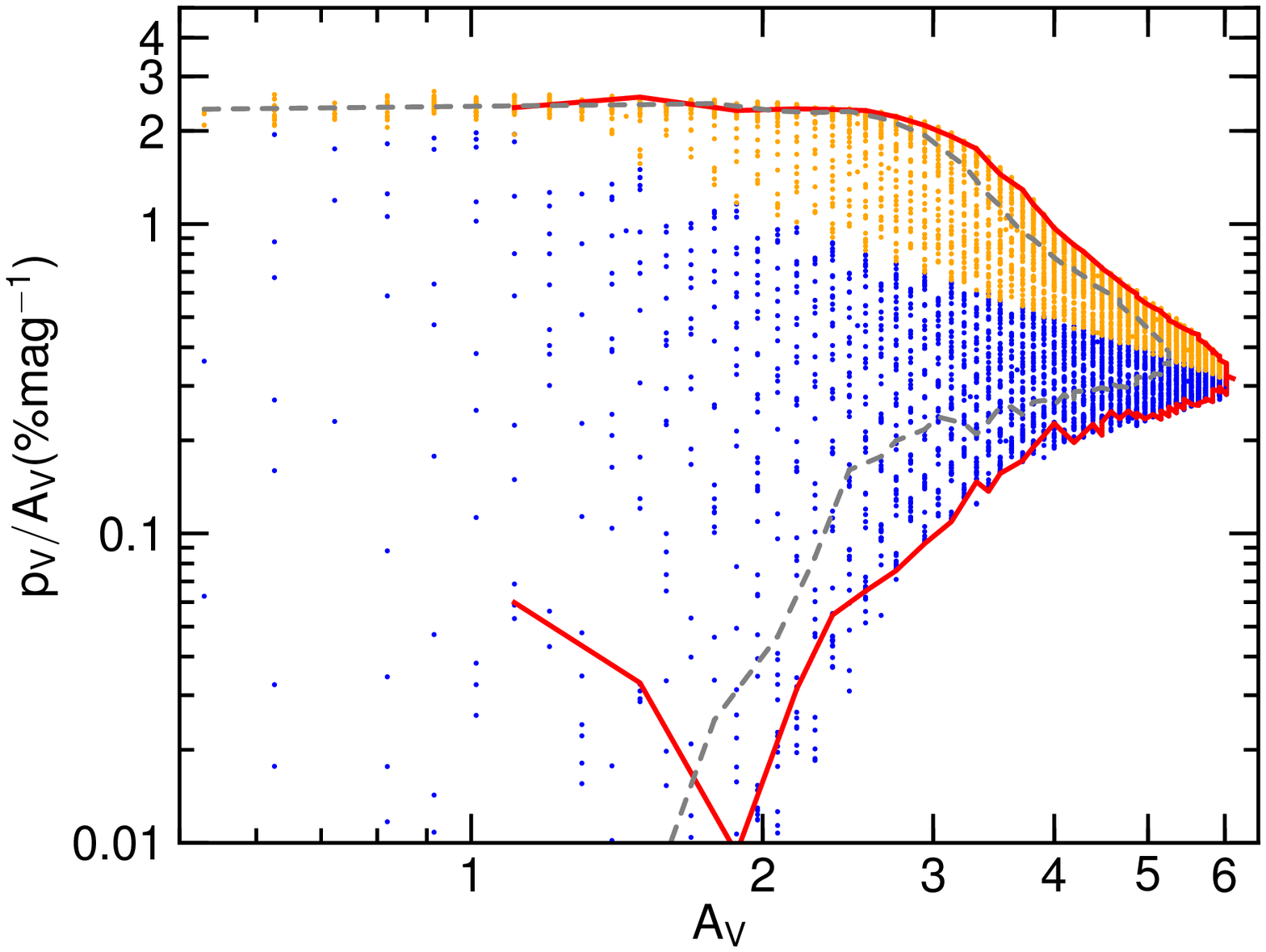}
\includegraphics[width=0.45\textwidth]{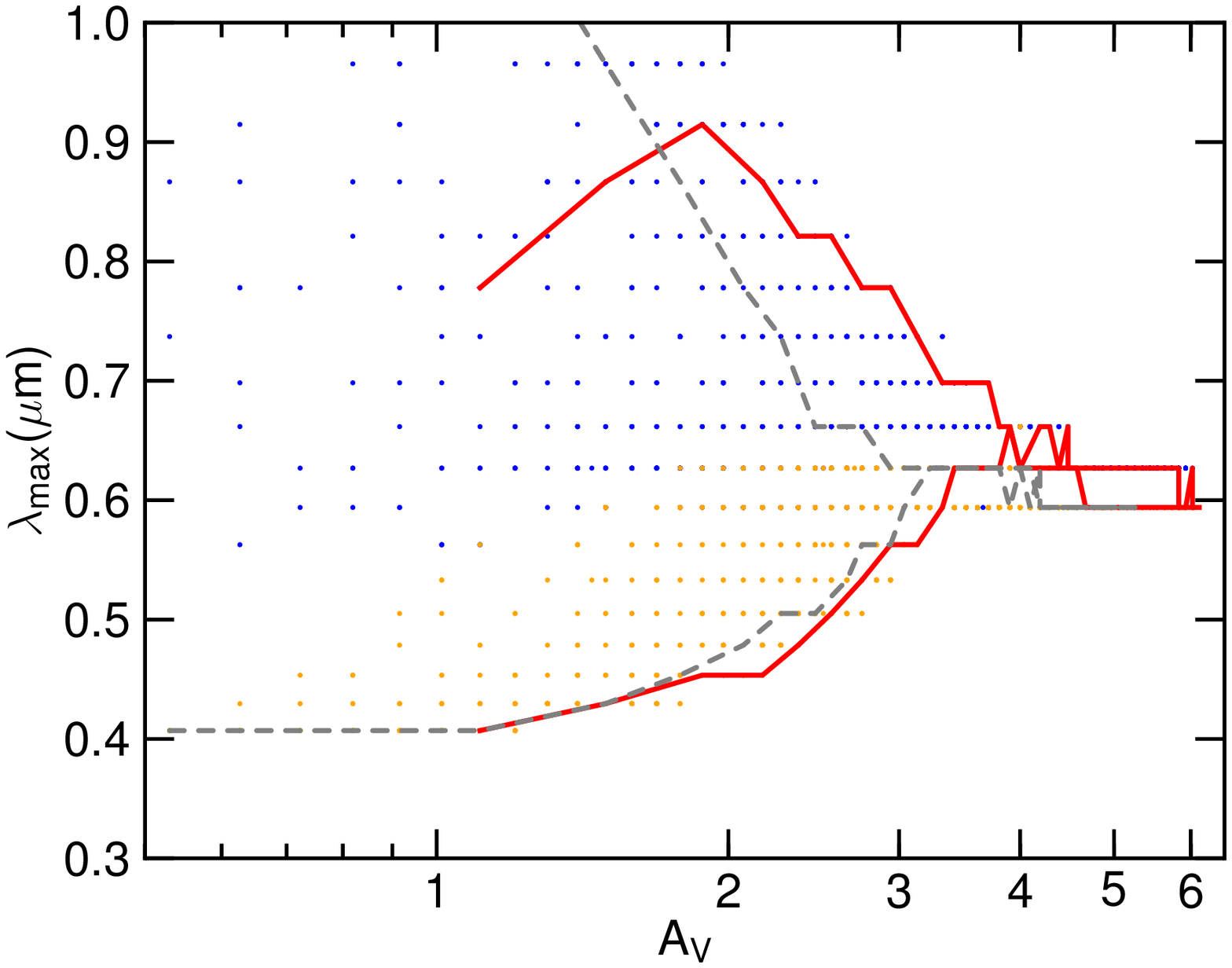}
\caption{Variation of $p_{V}/A_{V}$ (left) and $\lambda_{\max}$ (right) vs. $A_{V}$ over the entire IC 63 surface. Each dot corresponds to a sightline toward background stars. Orange circles denote the lines of sight through the dayside and blue circles denote the lines of sight through the nightside of IC 63. Solid and dashed lines show the variation $p_{V}/A_{V}$ along the cuts $Y/r_{c}=0$ and $Y/r_{c}=-0.5$. }
\label{fig:pV_full}
\end{figure*}

Figure \ref{fig:pV_full} shows our model predictions for $p_{V}/A_{V}$ (left) and $\lambda_{\max}$ (right) across the entire surface of IC 63. Results for the background stars behind the dayside (orange dots) exhibit higher $p_{V}/A_{V}$ and lower $\lambda_{\max}$ than the stars behind the nightside (blue dots). A sharp decline of $p_{V}/A_{V}$ at $A_{V}\sim 3$ is clearly seen, which is a direct consequence of the loss of grain alignment due to the reddening of radiation field (see also Figure \ref{fig:pmaxAV}).

\subsection{Grain Alignment by both RATs and H$_{2}$ formation torques}
Below, we investigate the effects of H$_{2}$ torques on the grain alignment and resulting polarization curves. 

The magnitude of H$_{2}$ torques is given by Equation (\ref{eq:JmaxH2}), which depends on the fraction of atomic hydrogen and the surface density of catalytic site. Calculations in \cite{1995A&A...302..223J} for IC 63 show that the fraction of atomic hydrogen decreases rapidly from the surface and becomes negligible ($\le 10^{-2}$) for the optical depth from the surface larger than $1$. Therefore, for our modeling, we account for H$_{2}$ torques for the outer region only, which is characterized by $\tau_{V,x}^{\gamma \rm Cas}\le 0.25$ where $\tau_{V,x}^{\gamma \rm Cas}=0.25$ corresponds to $\tau(\lambda=1100$\AA)=1 with $1110$\AA~ being the cut-off wavelength for H$_{2}$ photodissocitation. We assume the fraction of atomic hydrogen $n(\H)/n_{\H}=0.9$ for $\tau_{V,x}^{\gamma \rm Cas}\le 0.25$ and $n(\H)/n_{\H}=0.1$ for $\tau_{V,x}^{\gamma \rm Cas}>0.25$. To study the dependence of grain alignment on the magnitude of H$_{2}$ torques, we consider a grid of the surface density of active sites from $\alpha=10^{14}\cm^{-2}$ to $\alpha=10^{11}\cm^{-2}$. The rate of thermal flipping and the reduction factor $\Delta f$ as functions of grain size are taken from \cite{2009ApJ...695.1457H} (see Section 3.1 and 3.2 for discussion). The results presented here are for model 3, unless specified otherwise.

\begin{figure*}
\includegraphics[width=0.4\textwidth]{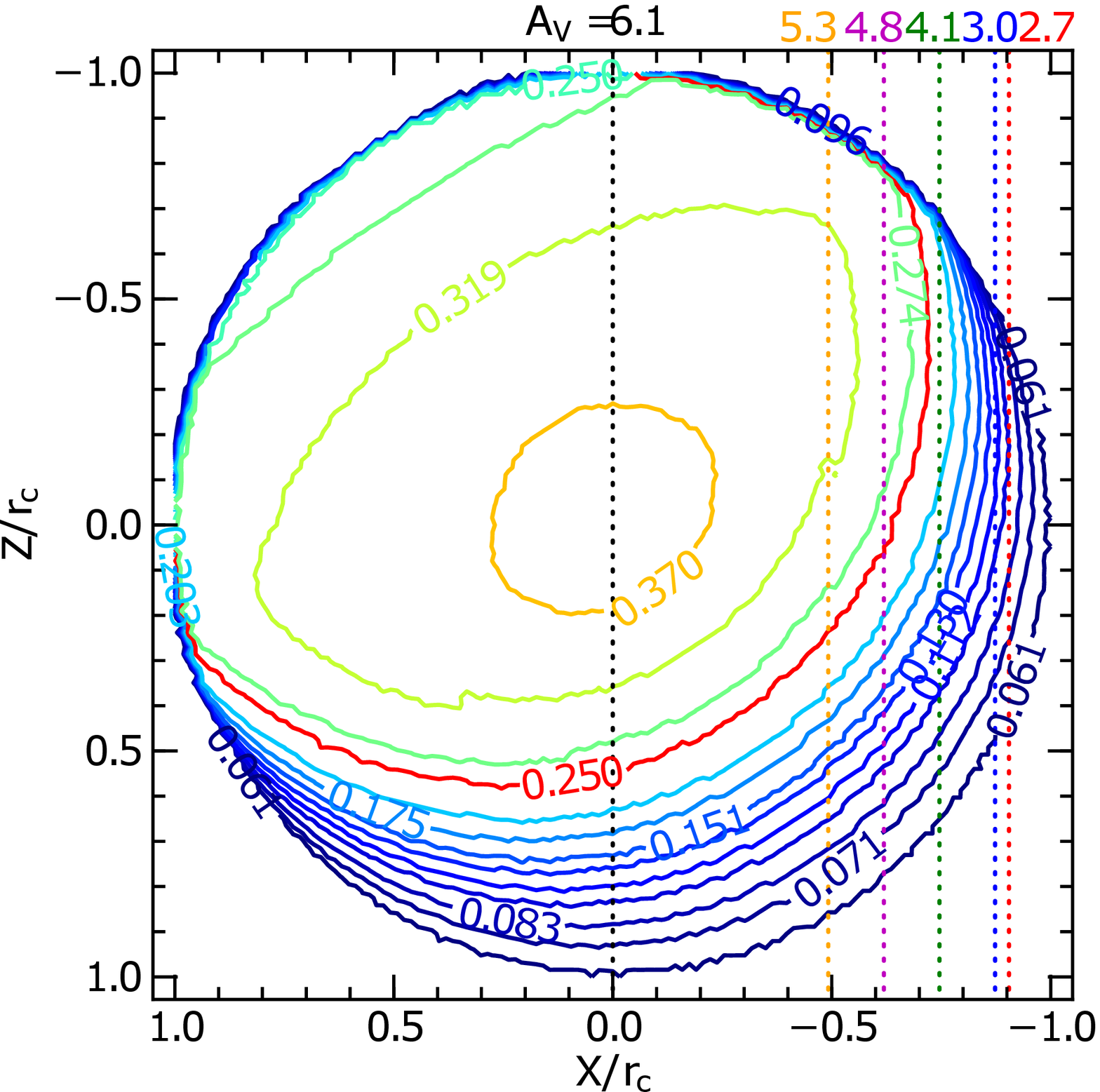}
\includegraphics[width=0.4\textwidth]{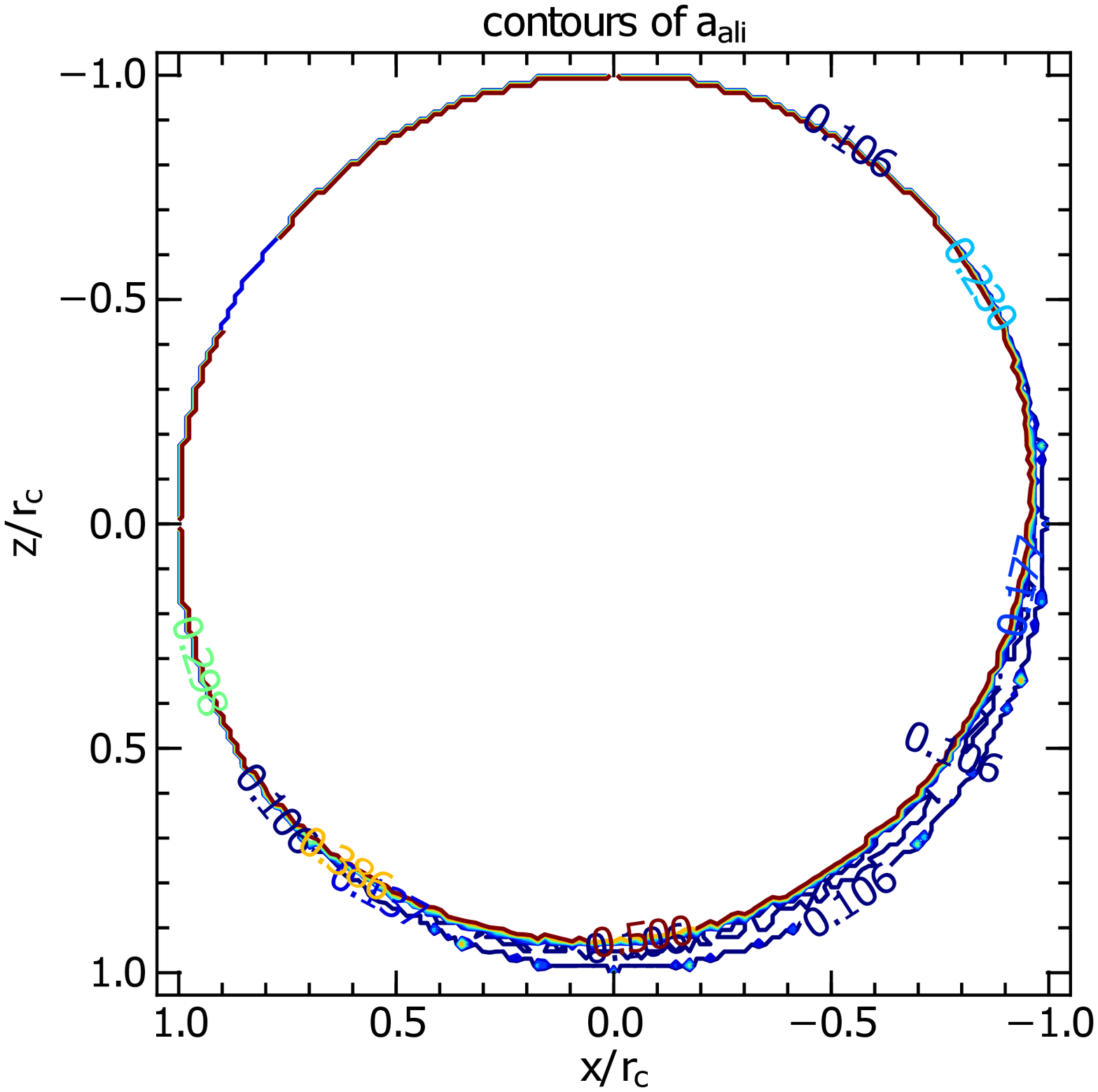}
\caption{Left panel: similar to Figure \ref{fig:aali_Tgas}, but for the alignment driven by both RATs and H$_{2}$ torques. Right panel: contours of grain size $a'_{\ali}$ above which grains are perfectly aligned due to effect of H$_{2}$ torques. The surface density of active site $\alpha=10^{12}\cm^{-2}$ is considered.}
\label{fig:aali_H2}
\end{figure*}

In the left panel of Figure \ref{fig:aali_H2} we show the contours of $a_{\ali}$ in the presence of both RATs and H$_2$ formation torques. Comparing to Figure \ref{fig:aali_Tgas}, we can see that the presence of H$_{2}$ torques has a minor effect on $a_{\ali}$, which indicates that, for the chosen $\alpha=10^{12}\cm^{-2}$, the value $a_{\ali}$ is mainly determined by strong RATs. The effect of H$_{2}$ torques in creating the new high-J attractor points is shown in the right panel where grains larger than $a'_{\ali}\sim 0.1\mum$ are assumed to be perfectly aligned.

\begin{figure}
\includegraphics[width=0.4\textwidth]{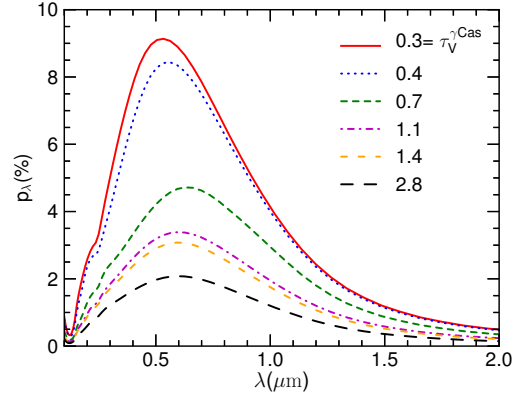}
\caption{Polarization curves for the alignment by both RATs and H$_{2}$ torques. The surface density of active site $\alpha=10^{12}\cm^{-2}$ is considered.}
\label{fig:plamH2}
\end{figure}

Figure \ref{fig:plamH2} shows the polarization curves for the RAT alignment with H$_2$ torques included. It can be seen that the polarization in general increases in the presence of H$_2$ torques. The effect is most significant for the sightlines with lower $\tau_{V}^{\gamma \rm Cas}$, which go through a thin layer near the surface. 

\begin{figure*}
\includegraphics[width=0.45\textwidth]{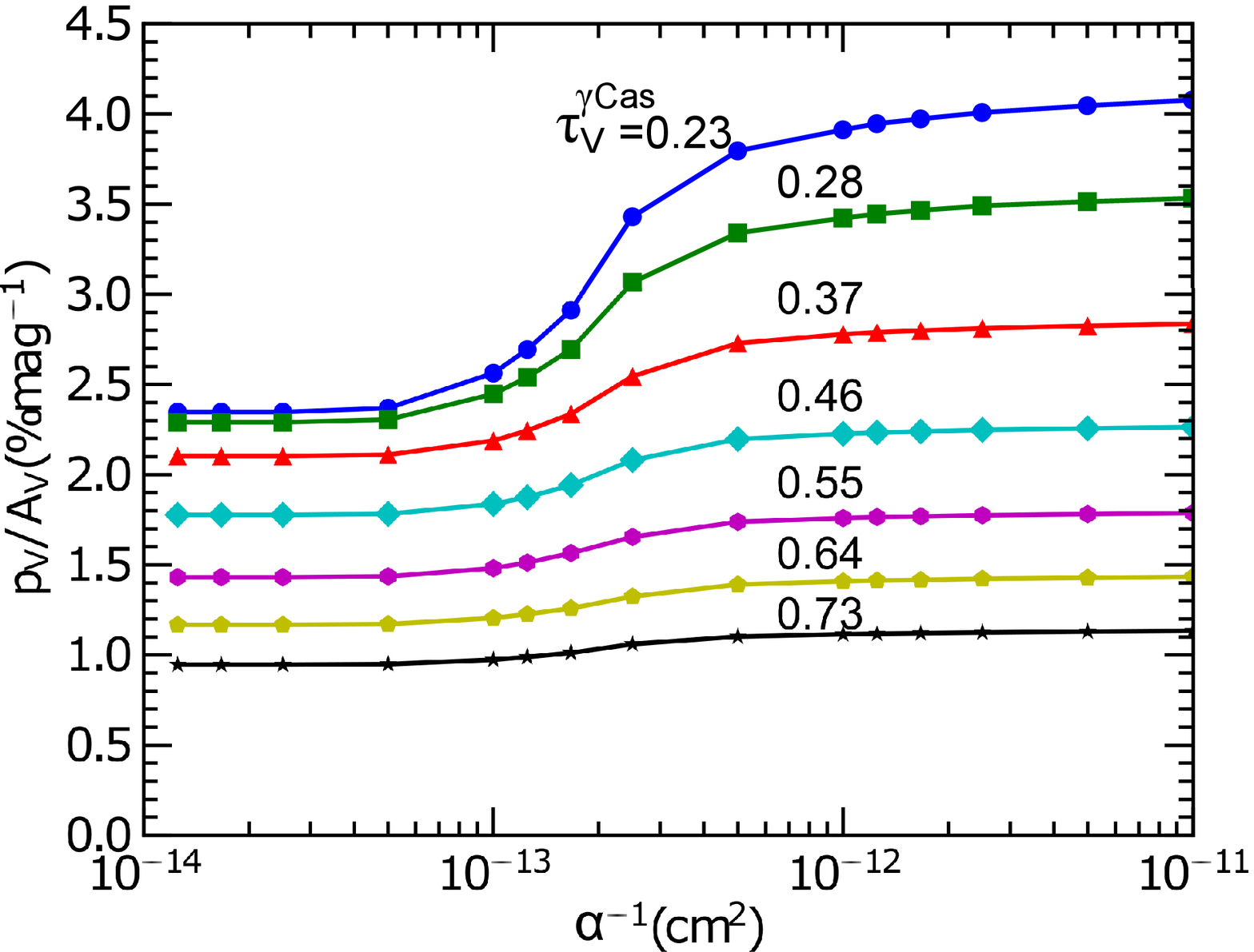}
\includegraphics[width=0.45\textwidth]{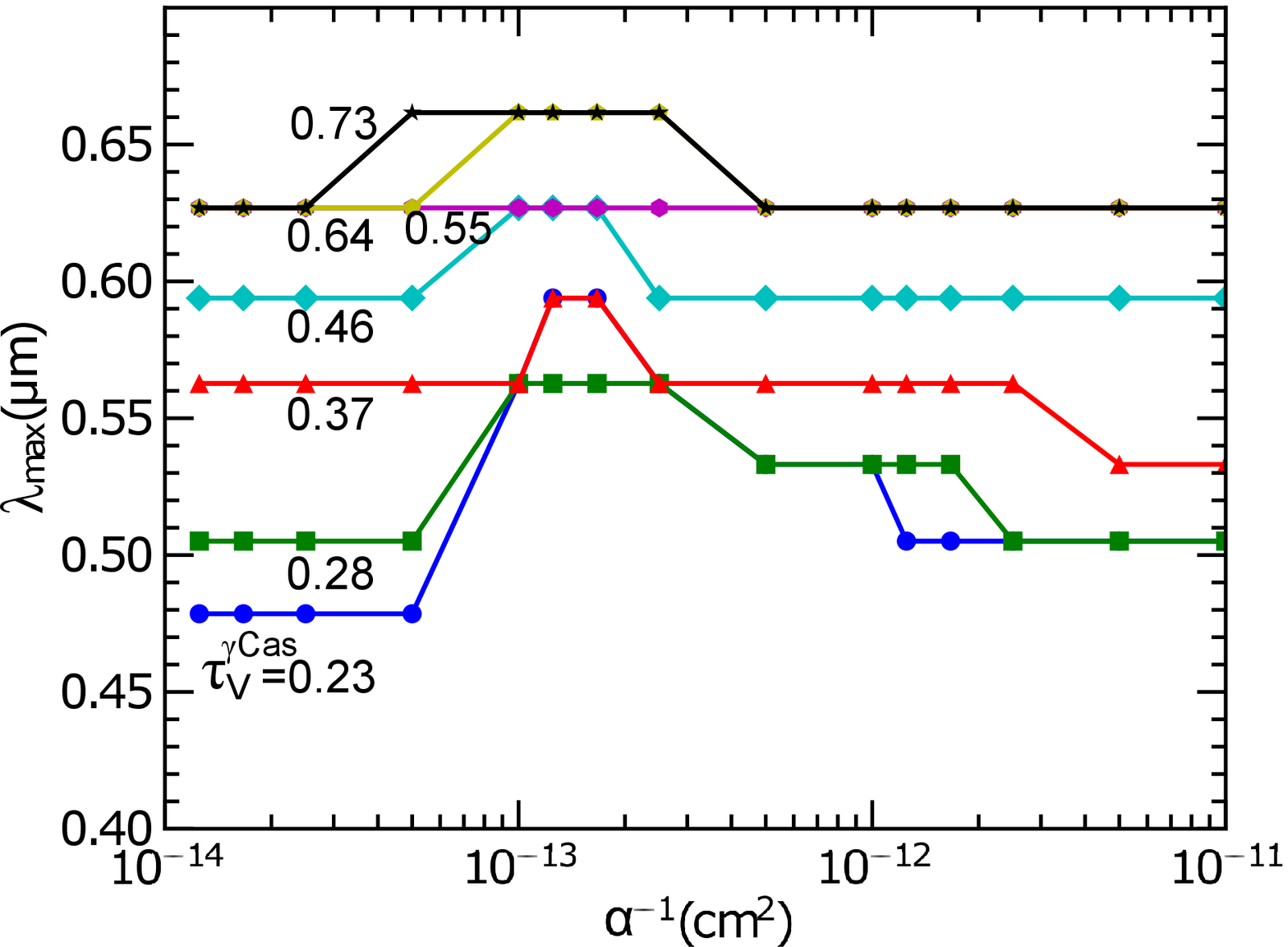}
\caption{Variation of $p_{V}/A_{V}$ (left) and $\lambda_{\max}$ (right) with the magnitude of H$_2$ torques ($\alpha^{-1}$) for different sightlines with $\tau_{V}^{\gamma \rm Cas}$ indicated. $p_{V}/A_{V}$ begins to increase with $\alpha^{-1}$ for $\alpha^{-1}> 5\times 10^{-14}\cm^{2}$. The middle slab ($Y/r_{c}=0$) is considered.}
\label{fig:pVAV_alpha}
\end{figure*}

Figure \ref{fig:pVAV_alpha} (left panel) shows the variation of $p_{V}/A_{V}$ with the magnitude of H$_2$ torques, which is obtained by varying $\alpha^{-1}$ while other parameters are kept unchanged. As shown, $p_{V}/A_{V}$ tends to increase with the increasing $\alpha^{-1}$ when the H$_2$ torques are sufficiently strong (i.e. $\alpha^{-1}> 5\times 10^{-14}\cm^{2}$). The effect is most profound for the sightlines through the thin surface layer with $\tau_{V}^{\gamma \rm Cas}<0.5$.

The variation of the peak wavelength $\lambda_{\max}$ with $\alpha^{-1}$ is shown in the right panel of Figure \ref{fig:pVAV_alpha}. It shows that $\lambda_{\max}$ does not increase monotonically with $\alpha^{-1}$. For sightlines with $\tau_{V}^{\gamma \rm Cas}<0.5$, $\lambda_{\max}$ increases with increasing $\alpha^{-1}$ initially, then it falls when $\alpha^{-1}$ becomes sufficiently large at $\alpha^{-1}\sim 10^{13}\cm^{2}$. Such a variation can be explained as the following. Due to strong RATs near the surface, even small grains can be radiatively aligned. For these small grains, the thermal flipping significantly suppresses the effect of H$_2$ torques acting at their high-$J$ attractor points, such that H$_{2}$ torques have minor effects on $a_{\ali}$ determined by RATs. However, at low-$J$ attractor points, H$_2$ torques can drive some intermediate ($a>a'_{\ali}$) grains to suprathermal rotation, increasing the degree of alignment of intermediate and large grains. As a result, the contribution of aligned, large grains to the total polarization increases, resulting in larger $\lambda_{\max}$. When H$_2$ torques become very strong (rather high $\alpha^{-1}$), the range of perfect alignment becomes broader due to the decrease of $a'_{\ali}$ (the upper dashed line in Figure \ref{fig:fhighJ} extends to rather small grains). As a result, $\lambda_{\max}$ is shifted to the blue. For the range of $\alpha$ considered and $\tau^{\gamma \rm Cas}_{V}<0.3$, $\lambda_{\max}$ is in general larger than that in the case without (or low) H$_2$ torques (see right panel).

\subsection{Comparison of model predictions with observations and evidence of $\H_{2}$ torques}

\begin{figure}
\centering
\includegraphics[width=0.45\textwidth]{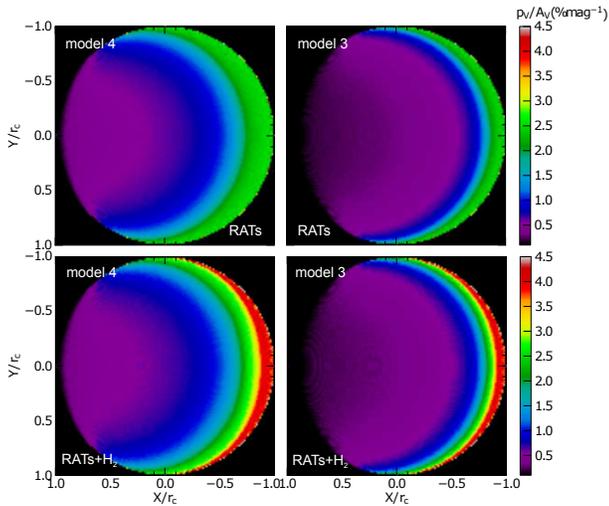}
\caption{Fractional polarization $p_{V}/A_{V}$ in the POS for the grain alignment by RATs without (upper) and with (lower) H$_{2}$ torques. The fractional polarization is strongest near the cloud surface facing $\gamma$ Cas. The presence of H$_{2}$ torques increases significantly the fractional polarization in the thin layers (red regions) of the dayside surface.}
\label{fig:image_PVAV}
\end{figure}

Figure \ref{fig:image_PVAV} shows the maps of $p_{V}/A_{V}$ predicted for grain alignment by RATs (upper panels) and by both RATs and H$_2$ torques (lower panels). Model 4 (left panels) and 3 (right panels) of IC 63 are shown. The presence of H$_{2}$ torques results in the increase of $p_{V}/A_{V}$ in a thin layer close to $\gamma$ Cas, extending from $Y/r_{c}\sim -1$ to $Y/r_{c}\sim 1$ (red region). Model with lower density appears to have larger $p_{V}/A_{V}$ at the same distance from $\gamma$ Cas (X value to the right).

\begin{figure}
\includegraphics[width=0.4\textwidth]{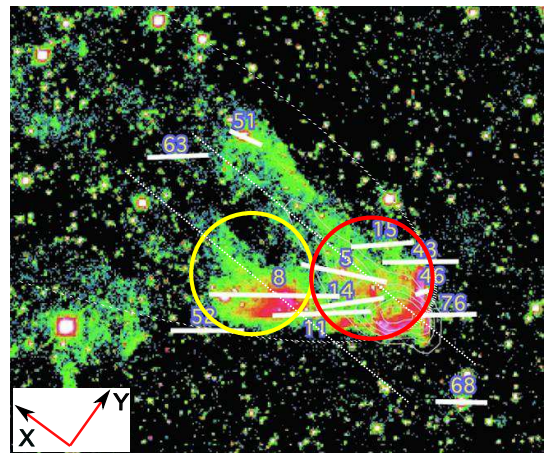}
\caption{Projections of the observed stars onto the POS with their observed V-band polarization vectors (white bars) are indicated. Image shows the intensity of H$_{2}$ fluorescence through a narrow-band filter centered on the H$_2$ 1--0 S(1) line at 2.122 $\mum$ for IC 63. Red circle indicates the clump at the tip of IC 63 that is adopted for our detailed modeling, and a nearby clump is marked by the yellow circle. Dotted lines are parallel to the direction of illumination by $\gamma$ Cas and go through the center of circles. Adapted from Andersson et al. (2013).}
\label{fig:IC63_tail}
\end{figure}

To facilitate comparison between predicted results and observational data, in Figure \ref{fig:IC63_tail}, we show the image of H$_{2}$ fluorescence and the polarization vectors observed for some background stars. Most sightlines to the background stars appear to probe the clump at the tip of IC 63 (red circle), whereas two sightlines to stars $\#$8 and $\#52$ probe a nearby clump (yellow circle). Moreover, stars $\#8$ and $\#11$ appear to lie behind the regions with the strongest H$_{2}$ fluorescence (i.e., strongest H$_{2}$ formation).\footnote{Multiwavelength observations in \cite{2005AJ....129..954K} show the complex structure of IC 63, with clumps and filaments.} We see some correlation between the region of strong H$_{2}$ fluorescence in the tip in Figure \ref{fig:IC63_tail} with the region of strongest $p_{V}/A_{V}$ in Figure \ref{fig:image_PVAV} (lower panels).

\begin{figure*}
\includegraphics[width=0.4\textwidth]{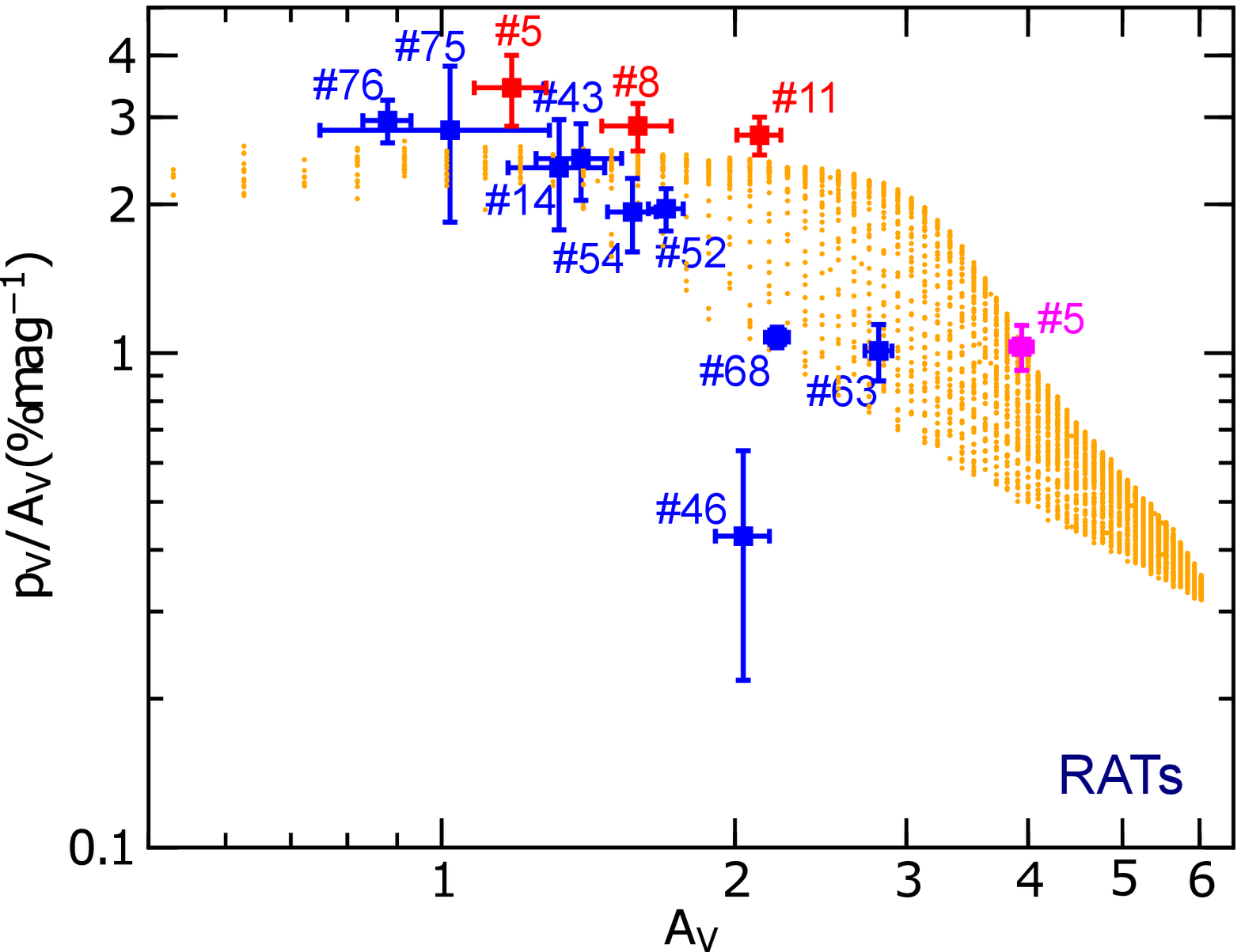}
\includegraphics[width=0.4\textwidth]{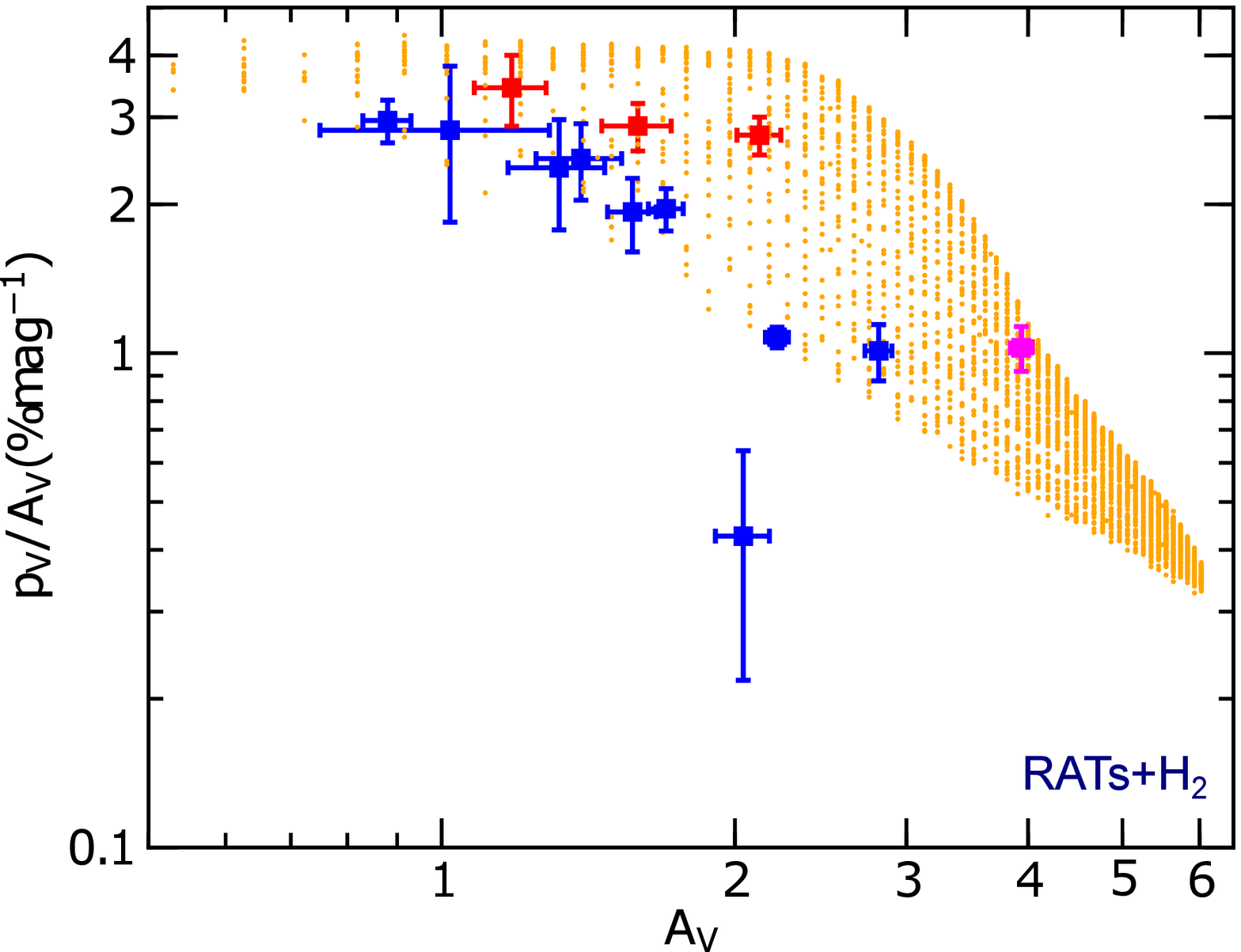}
\caption{Fractional polarization $p_{V}/A_{V}$ vs. $A_{V}$ from modeling (orange dots) compared with the observational data (filled square symbols) for the cases without (left) and with H$_{2}$ torques (right). Each dot corresponds to a sightline through IC 63. The observed stars with high $p_{V}/A_{V}$ are shown in red and other stars with lower $p_{V}/A_{V}$ are shown in blue. Red and purple colors for star $\#$5 indicate two possible $A_{V}$ estimated for this star. Predictions for background stars behind the dayside of IC 63 ($X< 0$) (dayside) are shown (see Figure \ref{fig:image_PVAV}).}
\label{fig:pVxcut_obs}
\end{figure*}

In Figure \ref{fig:pVxcut_obs} we plot $p_{V}/A_{V}$ versus $A_{V}$ obtained from modeling (orange dots) against the observational data (square symbols) from \cite{2013ApJ...775...84A}. Here model 3 is considered. Stars $\#5, \#8$ and $\#11$ with indication of stronger H$_{2}$ formation are shown in red squares and the remaining stars are shown in blue squares. \footnote{It is worthy noting that for star \#$5$, with indication of strong H$_{2}$ formation rate, the $A_{V}$ value is uncertain. The observationally derived value of the total-to-selective extinction (R$_V$) for this star is very large relative to the rest of the sample. \cite{2013ApJ...775...84A} argued that this might be due either to that this sightline contains unusually large grains, in which case the estimated value of $A_{V}=4.06$ (marked with purple) is appropriate.  If, on the other hand, the large R$_V$ is due to, e.g., a unresolved red companion (i.e. a systematic photometric error), calculating $A_{V}$ with an R$_V$ value based on the average of the remaining background stars would yield $A_{V}=1.18$.} 

Figure \ref{fig:pVxcut_obs} (left panel) indicates that the RAT alignment can successfully reproduce the observational data for the stars with weak H$_2$ formation. However, the alignment by only RATs cannot reproduce the highest values $p_{V}/A_{V}$ for the three stars ($\#5, \#8, \#11$) having strong H$_2$ formation. In the presence of H$_{2}$ torques, the right panel shows that our model can successfully reproduce the highest $p_{V}/A_{V}$ of these stars. {\it Is this a potential evidence for the enhancement of alignment by H$_2$ torques?} The answer is probably yes. 

From the left panel, it is apparent that the model of RAT alignment would reproduce the highest observed level if $p_V/A_V$ is increased by a factor $\sim 1.5$. This can be fulfilled by increasing the fraction of grains aligned with high-$J$ attractor points from $f_{\hi}=0.5$ to $f_{\hi}\sim 0.75$ (i.e., most of grains are perfectly aligned at high-$J$ attractor points). Based on theoretical considerations, such a high value $f_{\hi}$ seems less likely for the RAT alignment of normal paramagnetic dust and typical magnetic field strengths. Therefore, the effect of H$_2$ torques appears to be favored for the cause of enhanced $p_{V}/A_{V}$ observed for the stars with strong H$_2$ formation because it does not require a significant fraction of grains aligned at high-$J$ attractor points.

Let us compare the predicted results with the observational data in more detail. First, star $\#46$ has fractional polarization much lower than predicted by our model with $f_{\hi}=0.5$ (see Figure \ref{fig:pVxcut_obs}). This discrepancy stems from the fact that the sightline to this star probes the compression ridge of IC 63 having rather high gas density and temperature, which corresponds to enhanced collisional disalignment (see an extended discussion on this issue in \citealt{2013ApJ...775...84A}). 

Second, the model predictions are in good general agreement with the observation for the stars without (or low efficiency) H$_2$ formation. Indeed, the polarization of these stars (blue squares) can be reproduced by the model of RAT alignment, although two stars $\#75$ and $\#76$ are slightly above the prediction. Moreover, the photometry image shows that two stars $\#63$ and $\#68$ with low $p_V/A_V$ are located more distant from $\gamma$ Cas than the others. The polarization for these stars can be successfully reproduced by our modeling for sightlines through the middle region (dots in the lower bound).

The polarization of the stars with indication of efficient H$_{2}$ formation (red squares) can be reproduced by the model with H$_2$ torques evaluated near the surface closest to $\gamma$ Cas (dots in the upper bound in Figure \ref{fig:pVxcut_obs}). The sightline toward star $\#8$ appears to lie outside the spherical cloud (namely cloud 1) of the nebula that we carried out the modeling (red circle in Figure \ref{fig:IC63_tail}), but it can be considered to be in another spherical cloud (namely cloud 2, yellow circle in Figure \ref{fig:IC63_tail}). Assuming that the density of these two clouds are the same and the same radius, then, we expect that the grain alignment in cloud 2 is similar to that in cloud 1 for which these stars could be reproduced by the models for the stars near the surface of cloud 1.

Finally, the model with H$_2$ torques can also reproduce the fractional polarization for stars $\#75$ and $\#76$. Thus, these stars may have weaker collisional disalignment or have some alignment enhancement due to H$_2$ formation. 

\begin{figure*}
\includegraphics[width=0.4\textwidth]{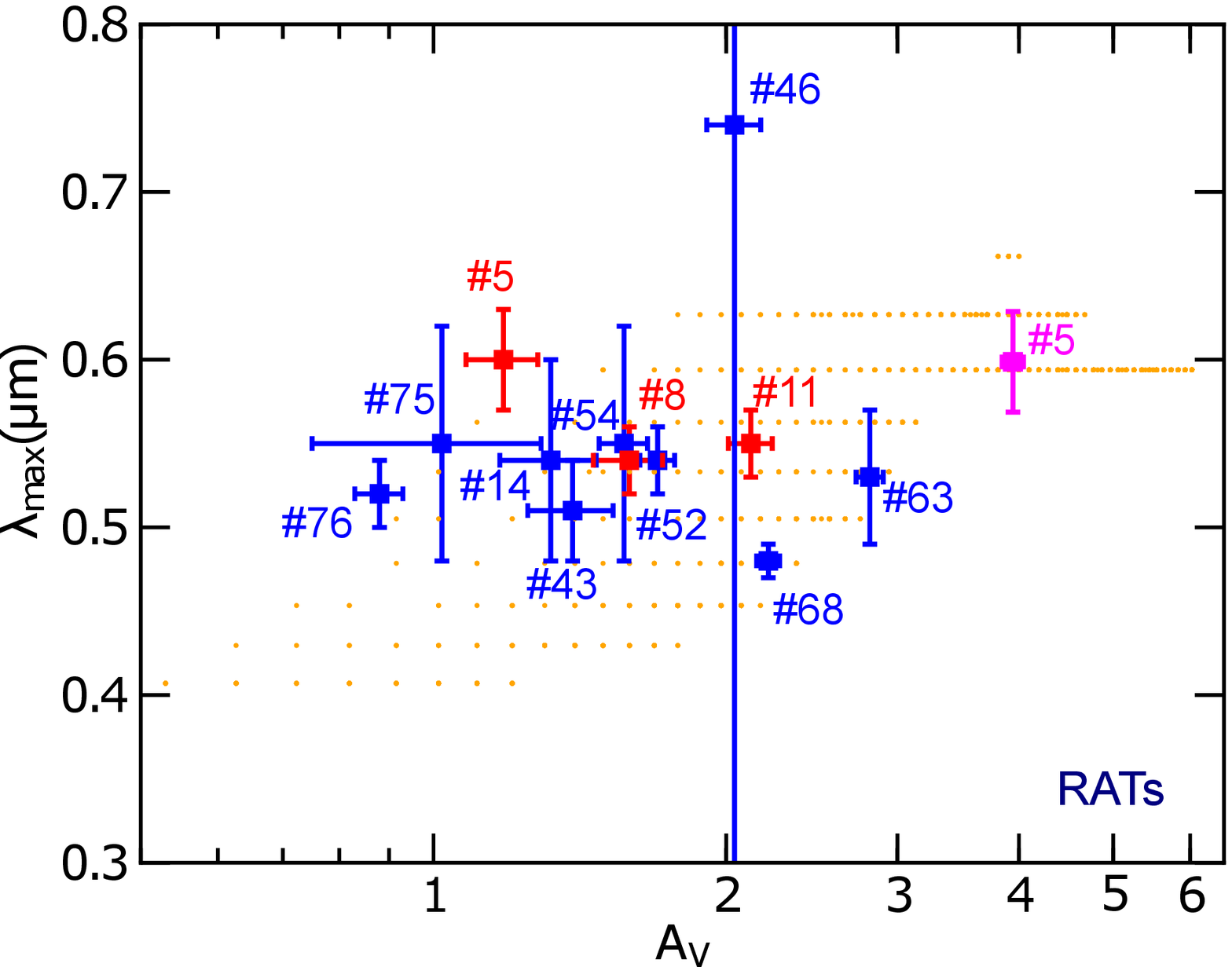}
\includegraphics[width=0.4\textwidth]{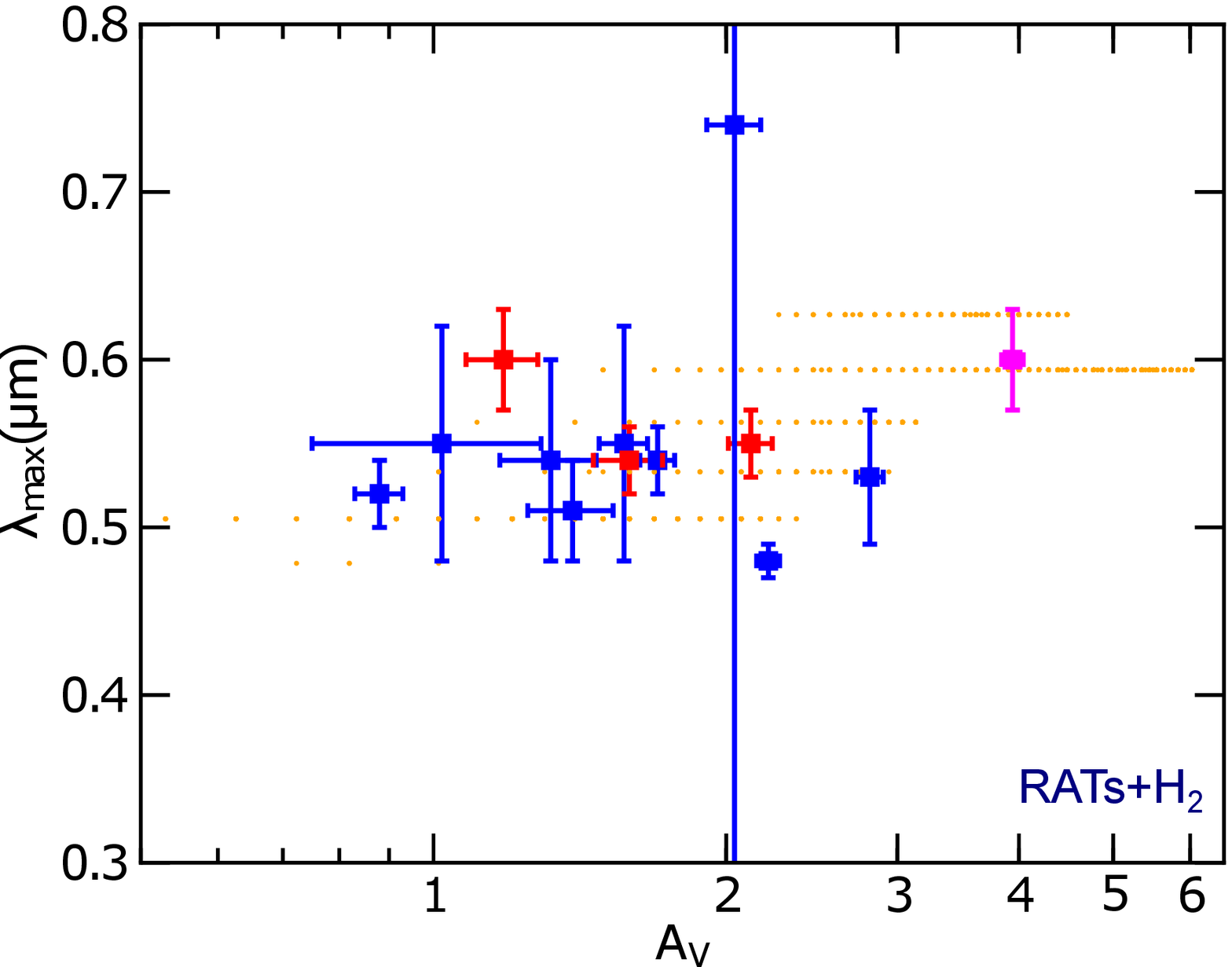}
\caption{Peak wavelength $\lambda_{\max}$ as a function of $A_{V}$. Left panel: predicted $\lambda_{\max}$ for RAT alignment from the dayside (orange dots) are below the observational values for $A_{V}<2$. Right panel: results with H$_2$ torques included, model predictions of $\lambda_{\max}$ (e.g. for $A_{V}<2$) become better agreement with the observations.}
\label{fig:lammax}
\end{figure*}

Figure \ref{fig:lammax} shows $\lambda_{\max}$ vs. $A_{V}$ for the case of alignment without (left) and with (right) H$_{2}$ torques. Near the surface (e.g., $A_{\V}< 2$), the alignment by only RATs has low $\lambda_{\max}$, which appears below the observed data (left panel). The inclusion of H$_2$ torques results in the increase of $\lambda_{\max}$ (see also Figure \ref{fig:pVAV_alpha}, right panel) in this outer layer and produces a better agreement between the predicted $\lambda_{\max}$ and the observational data (right panel).

The above comparison is based on our model 3, but we expect the main conclusions to remain valid also for model 1, 2 and 4, because of the systematic dependence of $p_{V}/A_{V}$ on $A_{V}$ (i.e., the gas column density).

\section{Discussion}\label{sec:dis}
\subsection{Previous studies on the RAT alignment in dense molecular clouds}

Grain alignment by RATs in starless, dense clouds has been studied by a number of authors (\citealt{2005ApJ...631..361C}; \citealt{2007ApJ...663.1055B}; \citealt{2008ApJ...674..304W}). These studies dealt with the alignment of grains by RATs induced by the attenuated ISRF. In this case, the alignment of grains inside the cloud depends on the radiation intensity determined by the depth (i.e., extinction) from the cloud surface and grain rotational damping (i.e., collisions through the gas density and temperature). Those studies concluded that grains can efficiently be aligned by RATs. Even in the cloud interior with $A_{V}\ge 10$, large grains can still be aligned because the RATs associated with the remaining, reddened light couple to the larger grain sizes. The fractional polarization is observed to decrease with increasing $A_{V}$, which is consistent with the predictions by RAT alignment mechanism.

{The simple, one-dimensional modeling of RAT alignment for a starless core in \cite{2008ApJ...674..304W} shows that the alignment in the dense cloud follows the model in which aligned grains are distributed in the outer layers, and the innermost region contains nonaligned grains. As a result, the fractional polarization is predicted to first decrease slowly with $A_{V}$ and then decline more rapidly according to a power-law $A_{V}^{-1}$ from some large $A_{V}\sim 3$.}

\subsection{RAT alignment in reflection nebula}
The present work is the first theoretical study dealing with the alignment of dust grains in a reflection nebula, a dense molecular cloud illuminated by a nearby star as well as the attenuated ISRF. For the conditions of high gas density and temperature of the reflection nebula, we found that the alignment is mainly determined by the strong stellar radiation, while the attenuated ISRF is inefficient in aligning grains, even in the thin layers of the envelope due to efficient collisional randomization. 

Our results also reveal that the fractional polarization is strong in the dayside facing the star and rather low in the nightside. In the spirit of RAT alignment, this is straightforward because grains closer to the surface would receive higher stellar radiation intensity, which results in the higher fractional polarization.

Observational studies for grain alignment in molecular clouds usually characterize the variation of the fractional polarization versus $A_{V}$ by a power-law ($p_{V}/A_{V}\propto A_{V}^{\eta}$) with a single slope, ranging from the cloud surface to center (e.g.,\citealt{2008ApJ...674..304W}; \citealt{2013ApJ...775...84A}). Based on detailed modeling of RAT alignment, it appears that such a description by a single slope may not reflect the realistic variation of grain alignment in the cloud. For IC 63, our results show that the variation of $p_{V}/A_{V}$ with $A_{V}$ in general follows two stages, well characterized by a power-law $p_{V}/A_{V}\propto A_{V}^{\eta}$ but with different slopes $\eta$. In the first stage, $p_{V}/A_{V}$ decreases slowly with $A_{V}$ with $\eta\sim -0.1$. In the second stage when the alignment of grains is significantly decreased, $p_{V}/A_{V}$ declines steeply with $\eta\sim -2$ (see Figure \ref{fig:pmaxAV}). 

The presence of a very steep slope ($\eta\sim -2$) starting from $A_{V}^{\rm tr}$ is surprising, although a steep slope $\eta\sim -1.1$ was already reported in \cite{2013ApJ...775...84A}. But it seems to be clear by considering the spherical geometry of IC 63 and its relative position with respect to the dominant radiation source ($\gamma$ Cas) in the framework of RAT alignment. Indeed, for the assumed spherical nebula, the alignment of grains inside the nebula is mainly produced by stellar radiation, which depends on the relative position of $\gamma$ Cas and IC 63. Grains in the dayside are directly illuminated by stellar radiation and can be aligned efficiently, whereas grains in the nightside receive much less radiation and are very weakly aligned. The outer layers containing aligned grains have a thickness that decreases with increasing distance from the star, corresponding to an increasing $A_{V}$ (see Figure \ref{fig:aali}, upper panel). Since the polarization of starlight is proportional to the column density of aligned grains along the sightline, the decrease in thickness of the layer with aligned grains results in the decrease of the polarization for the case of constant gas density assumed for IC 63 (see \ref{fig:aali}, lower panel). As a result, the fractional polarization $p/A_{V}$ declines steeper than $1/A_{V}$ predicted by the model of grain alignment with a constant polarization in the thin envelope of the cloud.

The transition visual extinction $A_{V}^{\rm tr}$ between the shallow and very steep slopes reveals the extent in the cloud where the RAT alignment is still significant. Because the RAT alignment is sensitive to grain size, observational determination of $A_{V}^{\rm tr}$ would provide useful insight into grain evolution in dense cores. We believe that the variation of $p/A_{V}$ versus $A_{V}$ following the different slopes is a generic feature of RAT alignment in clouds without embedded stars and illuminated by nearby stars similar to IC 63. Further observations for grain alignment in reflection nebulae would be useful to consolidate the RAT alignment theory. 

It is important to note that the observed fractional polarization will also be affected by the topology and turbulence of the magnetic field along the sightline \citep{1992ApJ...389..602J}, which is expected to result in a slope steeper than $\eta\sim -0.2$ for moderate $A_{V}$ predicted by us for uniform magnetic fields.

\subsection{Evidence for alignment enhancement by $\H_{2}$ torques}

\cite{2013ApJ...775...84A} found a potential correlation between the polarization for background stars and the intensity of H$_2$ fluorescence $I(\H_{2})$ from IC~63, while based on a relatively small number of points. They suggested that this correlation provides evidence for enhanced alignment of grains by H$_{2}$ torques. Based on our detailed modeling for grain alignment in a uniform spherical cloud, simulating IC 63, we find that the fractional polarization $p_{V}/A_{V}$ tends to increase with the increasing the magnitude of $\H_2$ torques, and that it is highest in the thin surface layer facing $\gamma$ Cas where H$_2$ formation takes place. Moreover, we find that the highest $p_{V}/A_{V}$ of the stars with indication of strong H$_2$ formation can successfully be reproduced when H$_{2}$ torques are incorporated within the RAT alignment paradigm.

We also showed that the model by RAT alignment cannot reproduce the highest $p_{V}/A_{V}$ observed when $50$ percent of grains aligned with high-$J$ attractor points (i.e., $f_{\hi}=0.5$). This problem can be resolved by increasing $f_{\hi}$ by a factor of $1.5$, i.e., from $f_{\hi}=0.5$ to $f_{\hi}\sim 0.75$. However, such a value $f_{\hi}=0.75$ perhaps is too high for the alignment by only RATs under typical conditions of the ISM. 

We find that the joint action of H$_2$ torques and RATs can successfully reproduce the observed highest $p_{V}/A_{V}$ with a more reasonable value $f_{\hi}=0.5$. It is noted that the increase $f_{\hi}$ to unity can be met by the inclusion of superparamagnetic material (iron clusters) into the grain. However, if this is true, then we may not see the increase of the polarization degree with the fluorescent emission intensity as observed in \cite{2013ApJ...775...84A} because the grains with superparamagnetic inclusions eventually have perfect alignment by RATs.

The variation of the peak wavelengths $\lambda_{\max}$ in the presence of H$_2$ torques appears to support the idea of alignment enhancement due to H$_2$ torques. Observational data reveals that the averaged peak wavelength in the case of strong H$_{2}$ formation is slightly larger than in the case without H$_2$ formation (\citealt{2013ApJ...775...84A}). Strong H$_2$ torques are found to have little effect on the alignment of small grains $a_{\ali}$ due to rapid flipping but can increase $f_{\hi}$ of large grains to unity (see Figure \ref{fig:aali_H2}). As a result, the peak polarization tends to shift to the red (see the right panel of Figure \ref{fig:pVAV_alpha}), and the predicted $\lambda_{\max}$ becomes in better agreement with observed ones (see the right panel of Figure \ref{fig:lammax}).

\subsection{Constraints on physical parameters of H$_2$ formation}
Our modeling predicts the increase of the fractional polarization $p_{V}/A_{V}$ with the increasing magnitude of $\H_{2}$ torques characterized by $\alpha^{-1}$ (see Figure \ref{fig:pVAV_alpha}). Interestingly, the observational data from \cite{2013ApJ...775...84A} reveal some correlation between $p_{V}/A_{V}$ versus the intensity of H$_2$ fluorescence, I$(\H_{2})$. Using this correlation, it may be possible to constrain the value $\alpha$ assuming a constant H$_2$ formation efficiency $\gamma_{\rm H}$ or vice versa. 

Based on observational data in (see \citealt{2013ApJ...775...84A}), we estimate the averaged ratio of $p_{V}/A_V$ for stars with high H$_2$ fluorescence to that of stars with zero H$_2$ fluorescence is $1.8\pm 0.4$. For the low H$_2$ fluorescence, the averaged ratio is $1.1\pm 0.1$. From Figure \ref{fig:pVAV_alpha} we see that the cases $\alpha^{-1}<5\times 10^{-14}\cm^{2}$ has a negligible effect on $p_{V}/A_{V}$ as well as $\lambda_{\max}$. Thus, the case $\alpha^{-1}=10^{-14}\cm^{2}$ can be considered as the case of without H$_2$ fluorescence. Therefore, the value $\alpha$ required to produce high H$_2$ fluorescence can be estimated as the following:
\bea
\frac{p_{AV}(\rm high~ \alpha^{-1})}{p_{AV}(\rm low~ \alpha^{-1})}=\frac{p_{AV}(\rm high ~I_{\H_{2}})}{p_{AV}(\rm zero~I_{\H_{2}})}=1.8,
\ena
where $p_{AV}= p_{V}/A_{V}$.

Using the sightline near the surface with $\tau_{V}^{\gamma \rm Cas}=0.23$ from Figure \ref{fig:pVAV_alpha}, one obtains
\bea
p_{AV}(\alpha^{-1})=1.8\times p_{AV}(\alpha^{-1}=10^{-14}\cm^{2})\approx 4.2,
\ena
which requires $\alpha^{-1}\sim 10^{-11}\cm^{2}$. 

For the case of low H$_2$ fluorescence, we get
\bea
p_{AV}(\alpha^{-1})= 1.1\times p_{AV}(\alpha^{-1}=10^{-14}\cm^{2})\approx 2.6,
\ena
which corresponds to $\alpha^{-1}\sim 10^{-13}\cm^{2}$.

As a result, the value $\alpha^{-1}$ is constrained in the range  $10^{-11}\cm^{2}>\alpha^{-1}> 10^{-14}\cm^{2}$ or $10^{14}\cm^{-2}>\alpha > 10^{11}\cm^{-2}$ assuming $\gamma_{\H}=0.1$.

Since the magnitude of H$_2$ torques is a function of $\gamma_{\H}\alpha^{-1/2}$ (see Equation \ref{eq:H2torq}). The value $\alpha$ would decrease if $\gamma_{\H}$ decreases. If we take the minimum $\alpha$ (strongest H$_2$ torque) corresponding to one active site per grain, then, we can also constrain the lower limit of $\gamma_{\H}$. For instance, for the $a=0.1\mum$ grain, we obtain $\alpha_{\min} \sim 10^{9}\cm^{-2}$, which yields $\gamma_{\H}> 0.1\times (10^{11}/10^{9})^{-1/2}>0.01$. 

\subsection{Recent observational studies of dust polarization}

{\it Planck} polarization data over the entire sky \citep{2014arXiv1405.0871P} shows a general decrease of the fractional polarization with gas column density $N_{\H}$. From their figure 18, we see that the variation of the mean polarization can be approximately described first by a shallow slope and then a steep slope $\eta=-1$ for $N_{\H}\ge 2\times 10^{22}\cm^{-2}$ (i.e., $A_{V}>10$), respectively. Our modeling of RAT alignment, although for a specific nebula (see Figure \ref{fig:pV_full}), seems to well reproduce the shallow slope observed. Moreover, the steep slope $\eta=-1$ for high $A_{V}$ seen by {\it Planck} is expected from the RAT alignment theory that predicts a significant loss of grain alignment toward the center of clouds.

\cite{Jones:2014vq} presented the variation of $p_{K}/\tau_{K}$ vs. $A_{V}$ for starless cores with $A_{V}$ up to $100$. They showed that the variation $p_{K}/\tau_{K}$ vs. $A_{V}$ can be fitted by a power-law with a shallow slope $\eta\sim -0.5$ for $A_{V}< 10$ and a steep slope starting from $A_{V}\sim 20$. While the effect of magnetic field turbulence is expected to reproduce the slope $\eta\sim -0.5$ for moderate $A_{V}$ (\citealt{1992ApJ...389..602J}, namely JDK model), it cannot explain the steep slope seen for high $A_{V}$. By including the decrease of grain alignment toward high $A_{V}$ as predicted by RAT alignment in the JDK model, Jones et al. successfully reproduced the slope $A_{V}^{-1}$ for $A_{V}>20$.

\cite{2014A&A...569L...1A} investigated the variation of $p/A_{V}$ for the starless Pipe nebula using optical, near-infrared (NIR), and submillimeter polarimetry. Interestingly, their NIR data show a steep slope $\eta\sim -1$ for $A_{V}\sim 5-9.5$ and a shallow slope $\eta \sim -0.34$ for $A_{V}\sim 9.5-30$. The steep slope is expected when the alignment of grains is significantly reduced according to the RAT alignment theory, whereas the shallow slope at higher $A_{V}$ is difficult to reconcile. The latter may originate from grain growth that occurs in very dense regions of the cloud \citep{2014A&A...569L...1A}, which results in additional alignment of big grains. It may also arise from magnetic field turbulence. \cite{1992ApJ...389..602J} showed that, for a magnetic field structure with a fixed mean magnetic field and a strong random component, when you transverse a few "unit cells" (low and moderate $A_{V}$), a slope of $\eta \sim {-0.5}$ is expected from a random walk effect. When the number of "cells" approaches infinity, the slope becomes shallower, because while the random field averages away, "eventually" the "constant" field shows back up again. 

Andersson et al. (2015) observed a clear break from a shallow slope $p/A_V \propto A_V^{-0.6}$ to steep slope $p/A_V \propto A_V^{-1.0}$ at $A_V\sim 20$ in the starless core L 183. These latest observational results indicate that the variation of the fractional polarization in molecular clouds is complicated and can be described by a power-law with the shallow and steep slopes. The RAT alignment naturally predicts multiple slopes for the variation of $p/A_{V}$ vs. $A_{V}$.

\subsection{Final notes}
Grain alignment stayed for decades as a mystery, which impeded the efforts to use dust polarization by aligned grains for studying magnetic fields. Shortly from the beginning of the research, the Davis-Greenstein alignment was identified as the major alignment process in spite of the problems with quantitative accounting for the observational data. The situation has been changed rather recently when the RAT alignment became a predictive theory. A big advantage of AMO is that it provides an analytical description of RATs that can account for the fundamental properties of RATs, and the form of the torques is simple enough that the effect of several physical processes in addition to RATs is possible to study. This has opened avenues for quantitative modeling of the RAT alignment. 

This paper is an attempt to provide the modeling of the effects of H$_2$ torques in addition to RATs for a particular astrophysical object in order to directly compare the theoretical predictions with observations. Lastly, with our aim to model the fundamental properties of RAT alignment theory, our present study does not consider explicitly the effects of the decrease of polarization due to magnetic turbulence (see \citealt{2005ApJ...631..361C}; \citealt{2007ApJ...663.1055B}). We expect our shallow slope for moderate $A_{V}$ will become steeper in the presence of magnetic turbulence. This issue will be addressed in detail in our future paper.

\section{Summary}\label{sec:sum}
We have carried out a detailed modeling of grain alignment and dust polarization for a reflection nebula using RAT alignment theory. The specific model includes a high, anisotropic, radiation intensity impinging on a cloud with a temperature characteristic of the diffuse ISM but a gas density typical of a dense molecular cloud. As such, the collisional damping rate of the grain alignment is expected to be significantly higher than in the diffuse ISM, which is also borne out by our modeling. Our principal results can be summarized as follows:

1. Grains are efficiently aligned by RATs in the outer layers of the dayside close to the star, whereas the alignment becomes less efficient in the nightside of the nebula due to the extinction of stellar radiation. In an ideal spherical cloud, the fractional polarization in the dayside facing the star is higher than in the nightside. 

2. For a middle slab in the direction of the stellar radiation, the fractional polarization $p_{V}/A_{V}$ slowly decreases with increasing $A_{V}$ (i.e., more distant from $\gamma$ Cas) and then rapidly declines from some transition value of $A_{V}$ due to the gradual loss of aligned grains in the cloud interior. The variation of $p_{V}/A_{V}$ with $A_{V}$ can be described by a power-law $A_{V}^{\eta}$ with a shallow slope $\eta=-0.1$ ($-0.2$) and very steep slope $\eta \sim -2$.

3. Our predictions for $p_{V}/A_{V}$ and peak wavelength $\lambda_{\max}$, based on the RAT alignment, are essentially consistent with the observational data. We show that the stars observed with lower fractional polarization (stars without indication of H$_2$ formation activity) can be reproduced by our model of alignment by RATs only.

4. We find the increase of the polarization fraction with the increasing magnitude of H$_2$ formation torques. Applying to IC 63, we show that the inclusion of H$_{2}$ formation torques in the model of RAT alignment can well reproduce the highest fractional polarization observed for stars with indication of strong H$_2$ formation. The agreement between the predicted and observed $\lambda_{\max}$ is greatly improved. These results provide valuable evidence in favor of a role for H$_{2}$ torques in grain alignment.  

5. Physical parameters related to H$_2$ formation, including the density of active site and H$_2$ formation efficiency, may be constrained using the observed fractional polarization and the model prediction. Additional data and more detailed three-dimensional modeling will allow this effect to be more definitely probed using our theoretical tools.

\section*{Acknowledgments}

We thank the referee for her/his insightful and valuable comments that improved our paper. A.L. acknowledge the financial support of the NSF grant AST-1109295, Vilas Award and the Center for Magnetic Self-Organization.  B-G A. acknowledge financial support from the NSF through grant AST-1109469. T.H. was supported by Humboldt Fellowship at Ruhr-Universit$\ddot{\rm a}$t Bochum.

\appendix

\section{Angle dependence of maximum angular momentim spun-up by RATs}\label{apdx:Jmax-psi}
Following AMO, the components of RAT efficiency after averaging over fast grain rotation can be approximated by (Equations (61) and (62) in LH07)
\bea
Q_{e1}(\Theta) &=& Q_{e1}^{\max}\left(5\cos^{2}\Theta-2 \right)/3,\\
Q_{e2}(\Theta) &=& Q_{e2}^{\max}\sin(2\Theta),
\ena
where $\Theta$ denotes the angle between $\hat{\ba}_{1}$ and $\kv$, and $Q_{ei}^{\max}$ is the maximum value of RAT component $Q_{ei}$. The $Q^{\max}$-ratio is simply defined as $q^{\max}=Q_{e2}^{\max}/Q_{e2}^{\max}$.

For the perfect coupling of $\hat{\ba}_{1}$ with $\bJ$ (perfect internal alignment), the RAT efficiency component parallel to $\bJ$ which acts to spin-up grains is defined as
\bea
 H=Q_{e1}(\Theta)\cos\Theta+Q_{e2}(\Theta)\sin\Theta,\label{eq:Hspin}
\ena 
and the projection of $\bGamma$ on to $\bJ$ is $\Gamma_{J}\propto H$.

Let consider an ambient magnetic field that makes an angle $\psi$ with $\kv$ and the perfect alignment of $\hat{\ba}_{1}, \bJ$ with $\Bv$. For this case, $\Theta\equiv \psi$, and Equation (\ref{eq:Hspin}) becomes
\bea
H = \left(Q_{e1}(\psi)\cos\psi+Q_{e2}(\psi)\sin\psi\right)
\ena

Plugging $Q_{e1}$ and $Q_{e2}$ into the above equation, we obtain
\bea
H = \left[ \frac{\left(5\cos^{2}\psi-2\right)q^{\max}}{3} +2\sin^{2}\psi\right]Q_{e2}^{\max} \cos\psi.\label{eq:H-spin}
\ena

The default model of AMO in which the inclination angle of the mirror is $45$ degree has $q^{\max}=1.2$ (see LH07). With this default $q^{\max}$ value, the term in the bracket of Equation (\ref{eq:H-spin}) becomes independent on $\psi$, and can be rewritten as
\bea
H=Q_{e1}(\psi=0)\cos\psi.
\ena

Since $\Gamma_{J}(\psi) \propto H= Q_{e1}(\psi=0)\cos\psi$, following Equation (\ref{eq:Jmax}) we can write
\bea
J_{\max}^{\RAT}(\psi)=J_{\max}^{\RAT}(\psi=0)\cos\psi.\label{eq:Jmax_psi}
\ena
\section{Derivation of long-lived H$_{2}$ torques}\label{sec:apdxH2}
Torques produced by hydrogen formations are studied by numerous authors. Here, we provide the main results for reference.

Let us consider a right square prism of width $2a_{2}$ and height $2a_{1}$ and $r=a_{2}/2a_{1}$. The total surface area is $S= 8r(r+1)(2a_{1})^{2}$. Let $\alpha$ be the density of active sites on the grain surface, thus, the total number of active sites is $\nu=\alpha S= \alpha 8r(r+1)(2a_{1})^{2}$. Note that each site has a radius of about 10 Angstrom, so numerous H atoms can be stuck and form simultaneously hydrogen molecules. Since each site has small area, the temperature of the site can be uniform, such that H molecules evaporate from the site can have the same mean kinetic energy but different direction. The torque induced by evaporating molecules from each active site can be calculated as follows.

Consider a surface perpendicular to $\hat{\ba}_{1}$ axis, the collision rate of H atoms from gas of density $dn(\bv)$ with velocity $\bv$ in $\bv,\bv+d\bv$ is given by
\bea
dR_{\coll}=dn(\bv)v_{z}dA, 
\ena
where $v_{z}>0$ is the component of velocity parallel to $\hat{\ba}_{1}$ along which the gas atoms collide with the grain surface, $dA$ is the surface element with the normal vector along $\hat{\ba}_{1}$, and
\bea
dn(\bv)=n_{\H}Z\exp\left(-\alpha v^{2}\right) dv_{x}dv_{y}dv_{z},
\ena
where $\alpha=m_{\H}/2k_{\B}T_{\gas}$ and $Z=(\pi/\alpha)^{-3/2}$ is the normalization coefficient. 

Integrating over the surface area and over isotropic distribution of gas atom velocity, we obtain
\bea
R_{\coll}&=&A_{1}\int v_{z}n_{\H}Z\exp\left(-\alpha v^{2}\right) dv_{x}dv_{y}dv_{z},\\
&=&A_{1}n_{\H}(\pi/\alpha)^{-3/2}(\pi/\alpha) \int_{0}^{\infty} v_{z}\exp\left(-\alpha v_{z}^{2}\right) dv_{z},\\
&=&A_{1}n_{\H}(\pi/\alpha)^{-1/2}(1/2\alpha)=A_{1}n_{\H}\frac{1}{2}\left(\frac{2k_{\B}T_{\gas}}{\pi m_{\H}} \right)^{1/2},\\
&=&A_{1}n_{\H}\frac{\langle v_{\H}\rangle}{4},
\ena
where $A_{1}$ is the surface area of the side $\hat{\ba}_{1}$, and $\langle v_{\H}\rangle=\left(8k_{\B}T_{\gas}/\pi m_{\H}\right)^{1/2}$ is the mean speed of gas atoms. Thus, the collision rate by gas atoms to any surface is equal to the surface area multiplied by the mean flux of incident atoms divided by four.

Due to its isotropic distribution of gas atoms, the total rate of H atoms arriving at the entire grain surface is given by 
\bea
\dot{N}_{\H}=S n_{\H}\frac{\langle v_{\H}\rangle}{4}=n_{\H}\langle v_{\H}\rangle 2r(r+1)(2a_{1})^{2},
\ena

Let $\gamma_{\H}$ be the fraction of arrival H atoms that form H$_{2}$ molecules, then, the total number of H$_{2}$ can be formed over the entire grain surface is
\bea
\dot{N}_{\rm mol}= \frac{\gamma_{H}}{2}\dot{N}_{\H}= \gamma_{\H}(1-y)n_{\H}\langle v_{\H}\rangle r(r+1)(2a_{1})^{2}.
\ena

The number of H$_{2}$ molecules formed per site is then $\dot{N}_{1}=\dot{N}_{\rm mol}/\nu$. Assuming that the grain is spinning around its symmetry axis, and we want to calculate the increase of grain angular momentum along its symmetry axis per units of time. As a result, only rockets from the sides of the grain can increase the grain angular momentum in the direction parallel to the grain symmetry axis (i.e., spin the grain up). The number of active sites on the sides is
\bea
N_{\rm side}= \alpha 4(2a_{1})\times (2a_{2})=\alpha 8r (2a_{1})^{2}.
\ena
Thus, the fraction of sites useful for spin-up is then $f_{\rm spin-up}=N_{\rm side}/\nu=1/(r+1)$.

The angular momentum deposited by an H$_{2}$ molecule is equal to
\bea
\delta L_{z} = r (2m_{\H})v\sin\theta,
\ena
where $\theta$ is the angle between the ejection direction of the H$_{2}$ molecule and the radius vector $r$. 

Assuming that the active site is narrow and deep below the grain surface, such that all H$_{2}$ molecules from one active site leave the grain in a single direction and at the same speed. The direction and velocity of escaping H$_{2}$ molecules from different active sites is random, and its instantaneous velocities are given by Maxwellian distribution. The total angular momentum deposited to the grain by all H$_{2}$ molecules leaving active site $i$:
\bea
\delta \mathcal{L}_{i}= \frac{\dot{N}_{\rm mol}}{\nu} r (2m_{\H})v \sin\beta,
 \ena
where $\beta$ is the angle between $\bv$ and $\rv$. Approximating the brick as a thin disk, then, $\beta=\theta$, the angle between $\bv$ and the grain tangential direction, and $r\approx a_{2}$.

Summing $\mathcal{L}_{i}$ over all active sites (summing over all escaping angles $\beta$) results in zero net angular momentum. However, the total increase of squared angular momentum per units of time is not averaged out to zero and takes the following form:
\bea
\Delta L_{z}^{2} = \sum_{i=1}^{\nu/(r+1)}\delta \mathcal{L}_{i}^{2}.
\ena

Because the velocity and direction of escaping molecules from the different active sites are random, we can average the above equation as follows:
\bea
\langle \Delta L_{z}^{2}\rangle &=&\frac{\nu}{r+1}\langle \delta \mathcal{L}^{2}\rangle,
\ena
where 
\bea
\langle \delta \mathcal{L}^{2}\rangle=\frac{\dot{N}_{\rm mol}^{2}}{\nu^{2}} r^{2} (2m_{\H})^{2}\int_{0}^{\infty}dv Z v^{2} e^{-\alpha v^{2}}v^{2} \int_{0}^{\pi}d\phi\int_{0}^{\pi}d\theta \sin\theta \cos^{2}\theta,
\ena
where the integration over $\phi$ is taken from $0$ to $\pi$ only due to the fact that only outward H$_{2}$ molecules contribute to the recoil.

Using the integral $\int_{0}^{\infty} e^{-\alpha v^{2}}dv=\pi^{1/2}/2\alpha^{-1/2}$ and $\int_{0}^{\infty} v^{4}e^{-\alpha v^{2}}dv=3\pi^{1/2}/8\alpha^{-5/2}$, we obtain
\bea
\langle \delta \mathcal{L}^{2}\rangle = \frac{\dot{N}_{\rm mol}^{2}}{\nu^{2}} a_{2}^{2}\frac{2m_{\H}E_{\kin}}{3},
\ena
where $E_{\kin}=3k_{\B}T_{\gas}/2$ is the mean kinetic energy of H$_{2}$ molecules.

Thus,
\bea
\langle \Delta L_{z}^{2}\rangle=\frac{\nu}{r+1}\frac{\dot{N}_{\rm mol}^{2}}{\nu^{2}}a_{2}^{2} \frac{2m_{\H}E_{\kin}}{3},
\ena

Plugging in $\dot{N}$ and $a_{2}=2a_{1} r$, we obtain
\bea
\langle \Delta L_{z}^{2}\rangle&=&r^{4}(r+1)\gamma_{H}^{2} (1-y)^{2} n_{\H}^{2}\langle v_{\H}\rangle^{2} (2a_{1})^{6} \left(\frac{2m_{\H}E_{\kin}}{3\nu}\right)
\ena

The magnitude of torque due to $\H_{2}$ formation is denoted as
\bea
\Gamma_{\H_{2}}\equiv \langle \Delta L_{z}^{2}\rangle^{1/2}.\label{eq:Gamma-H2}
\ena

\section{Extinction and Polarization cross-section}\label{apdx:Cext}
\subsection{Extinction cross-section}
Let us consider a spheroid grain with the symmetry axis $\hat{\ba}_{1}$. A perfectly polarized electromagnetic wave with the electric field vector $\bE$ propagates along the $z$-axis,
which is perpendicular to the symmetry axis. Let $C_{\ext}(\bE\perp \ba)$ and $C_{\ext}(\bE\| \ba)$ be the extinction of the radiation for the cases in which the electric field vector
is parallel and perpendicular to the grain symmetry axis, respectively.

For simplification, we denote these extinction cross-section by $C_{\perp}$ and $C_{\|}$.
For the general case in which $\bE$ makes an angle $\theta$ with the symmetry axis, the extinction cross-section becomes
\bea
C_{\ext}=\cos^{2}\theta C_{\|}+\sin^{2}\theta C_{\perp},\label{eq:Cext0}
\ena

Since the original starlight is unpolarized, one can compute the total extinction cross-section by integrating Eq. (\ref{eq:Cext0}) over the isotropic distribution of $\theta$, i.e., $f_{\rm iso} d\theta= \sin\theta/2 d\theta$. As a result,
\bea
C_{\ext}=\frac{1}{3}\left(2C_{\perp}+C_{\|}\right).\label{eq:Cext}
\ena

Throughout this paper, the polarization cross-section is defined as
\bea
C_{\pol}=C_{\perp}-C_{\|}, ~C_{\pol}=\frac{1}{2}\left(C_{\|}-C_{\perp}\right),
\ena
for oblate and prolate spheroidal grains, respectively.

\subsection{Polarization cross-section}
Consider an observation coordinate system, which is defined by the sightline directed along the $z$-axis, the projection of the magnetic field on the POS denoted by the $y$-axis, and the third axis is perpendicular to the $yz$ plane, namely $z$-axis. Thus, $\bB$ lies in the $yz$ plane and makes a so-called angle $\xi$ with the $y$-axis.

By transforming the grain coordinate system to the observer coordinate system and taking corresponding weights, we obtain
\bea
C_{x}&=&C_{\perp}-\frac{C_{\pol}}{2}\sin^{2}\beta,\\
C_{y}&=&C_{\perp}-\frac{C_{\pol}}{2}(2\cos^{2}\beta\cos^{2}\xi+\sin^{2}\beta\sin^{2}\xi),
\ena
where the perfect internal alignment of grain axes with the angular momentum has been assumed.

The polarization cross-section then becomes
\bea
C_{x}-C_{y}=C_{\pol}\frac{\left(3\cos^{2}\beta-1\right)}{2}\cos^{2}\xi.\label{eq:Cpol}
\ena
Taking the average of $C_{x}-C_{y}$ over the distribution of the alignment angle $\beta$, the above equation can be rewritten as
\bea
C_{x}-C_{y}=C_{\pol}\langle Q_{J}\rangle \cos^{2}\xi,\label{eq:Cx-Cy}
\ena
where
\bea
Q_{J}=\frac{\left(3\cos^{2}\beta-1\right)}{2}
\ena
is the degree of alignment of the grain angular momentum with the ambient magnetic field. 

When the internal alignment is not perfect, following the similar procedure, we obtain
\bea
C_{x}-C_{y}=C_{\pol}\langle Q_{J}Q_{X} \rangle \cos^{2}\xi \equiv
C_{\pol}R\cos^{2}\xi,\label{eq:Cpol}
\ena
where $R=\langle Q_{J}Q_{X}\rangle$ is the Rayleigh reduction factor.

\section{Radiative Transfer for Polarized Radiation}\label{sec:apdxa}
In general, polarized light can be described by the Stokes parameters, I, Q, U, and V.

\subsection{General Radiative Transfer Equations}

Consider the propagation of unpolarized starlight of intensity $I_{0}$ through a medium containing aligned grains. Let 
$\gamma_{x}, \gamma_{y}$ be the opacity in units of $\cm^{-1}$, when the electric field is parallel to the $\xhat$ and $\yhat$ directions, and $\psi$ is the angle between the projected magnetic field on to the POS and the $\xhat$ axis. Opacity is related to the cross-section per column density follows:
\bea
d\tau_{x}=\gamma_{x} dz=\sigma_{x}n_{\gas} dz,\\
d\tau_{y}=\gamma_{x} dz=\sigma_{y}n_{\gas} dz,
\ena
Thus,
\bea
\gamma_{x}=\int C_{x} \frac{dn}{da} da,~ \gamma_{y}=\int C_{y} \frac{dn}{da} da,\label{eq:gammaxy}
\ena
where $dn/da$ is the grain size distribution.

The opacity related to circular polarization are denoted by $\delta_{x}$ and $\delta_{y}$
As a result, RT equations can be written as
\bea
\frac{d\ln I}{dz}&=&-\frac{\gamma_{x}+\gamma_{y}}{2}+\frac{\Delta \gamma}{2}\left[\frac{Q}{I}\cos 2\psi+\frac{U}{I}\sin 2\psi\right],\\
\frac{d}{dz}\left(\frac{Q}{I}\right)&=&\frac{\Delta \gamma}{2}\cos 2\psi+\Delta \delta \frac{V}{I}\sin 2\psi\nonumber\\
&&-\frac{\Delta \gamma}{2}\left(\frac{Q}{I}\right)\left[\frac{Q}{I}\cos 2\psi+\frac{U}{I}\sin 2\psi\right],\\
\frac{d}{dz}\left(\frac{U}{I}\right)&=&\frac{\Delta \gamma}{2} \sin 2\psi-\Delta \delta \frac{V}{I}\cos 2\psi\nonumber\\
&&-\frac{\Delta \gamma}{2}\left(\frac{Q}{I}\right)\left[\frac{U}{I}\cos 2\psi+\frac{U}{I}\sin \psi \right],\\
\frac{d}{dz}\left(\frac{U}{I}\right)&=&\Delta \delta\left[\frac{U}{I}\cos 2 \psi-\frac{Q}{I}\sin 2\psi\right]\nonumber\\
&&-\frac{\Delta \delta}{2}\left(\frac{V}{I}\right)\left[\frac{Q}{I}\cos 2 \psi+\frac{U}{I}\sin 2\psi\right],
\ena
where $\Delta \gamma=\gamma_{x}-\gamma_{y}$ and $\Delta \delta=\delta_{x}-\delta_{y}$.

Denote $q=Q/I$, $u=U/I$ and $v=V/I$, the above equations become
\bea
\frac{d\ln I}{dz}&=&-\frac{\gamma_{x}+\gamma_{y}}{2}+\frac{\Delta \gamma}{2}\left[q\cos 2\psi+u\sin 2\psi\right],\\
\frac{dq}{dz}&=&\frac{\Delta \gamma}{2}\cos 2\psi+\Delta \delta v\sin 2\psi-\frac{\Delta \gamma}{2}q\left[q\cos 2\psi+u\sin 2\psi\right],\\
\frac{du}{dz}&=&\frac{\Delta \gamma}{2}\sin 2\psi-\Delta \delta v\cos 2\psi-\frac{\Delta \gamma}{2}u\left[q\cos 2\psi+u\sin 2\psi\right],\\
\frac{dv}{dz}&=&\Delta \delta\left[u\cos 2 \psi- q \sin 2\psi\right]-\frac{\Delta \gamma}{2} v\left[q\cos 2 \psi+u\sin 2\psi\right]
\ena

The polarization is then equal to
\bea
p=\sqrt{q^{2}+u^{2}}.\label{eq:linpol}
\ena

\subsection{Linear Polarized Light}
Since we are interested in the polarization by dichroic extinction from near UV to near infrared wavelength, we can disregard the emission term (second term in Equation of I) in radiative transfer (RT) equations. Moreover, by disregarding the second order terms of $q$ and $u$, we obtain
\bea
\frac{d\ln I}{dz}&=&-\frac{\gamma_{x}+\gamma_{y}}{2},\\
\frac{dq}{dz}&=&\frac{\Delta \gamma}{2}\cos 2\psi,\\
\frac{du}{dz}&=&\frac{\Delta \gamma}{2}\sin 2\psi,
\ena

Solving these equations, we obtain the Stokes parameters $q$ and $u$. The polarization is simply given by \ref{eq:linpol}.

For the simplified case in which the magnetic field uniform and does not change along the sightline, the polarization becomes
\bea
p\equiv \sqrt{q^{2}+u^{2}}=\int (dp/dz) dz,
\ena
where
\bea
\frac{dp}{dz}=\frac{\Delta \gamma}{2}=\frac{\gamma_{x}-\gamma_{y}}{2}.\label{eq:dpdz}
\ena
Using Equations (\ref{eq:gammaxy}), the above equation can be rewritten as
\bea
dp=\int_{a_{\min}}^{a_{\max}} \left(\frac{C_{x}-C_{y}}{2}\right)\frac{dn}{da}da dz=\int_{a_{\ali}}^{a_{\max}} \left(\frac{C_{x}-C_{y}}{2}\right)\frac{dn}{da}da dz,
\ena
where the assumption that grains smaller than $a_{\ali}$ are randomly oriented such that $C_{x}-C_{y}=0$. For $a>a_{\ali}$, the polarization cross-section $C_{x}-C_{y}$ is given in Equation (\ref{eq:Cpol}).

\section{Relation between $A_{V}$ and $\tau_{V}^{\gamma {\rm Cas}}$}
For a spherical cloud of radius $r_{c}$, the distance from the grain located at $(x,z)$ to the $\gamma$ Cas is described by
\bea
\Delta x_{s} = r_{c}-\sqrt{r_{c}^{2}-z^{2}}.
\ena

When the gas density is uniform, the optical depth relative to $\gamma$ Cas is obtained from the visual extinction to the background star as follows: 
\bea
1.086\tau_{V}^{\gamma \rm Cas}\equiv  A_{V}^{\max}/2-\left[(A_{V}^{\max}/2)^{2}-(A_{V}/2)^{2} \right]^{1/2},\label{eq:AV_to_tauV}
\ena
where $A_{V}^{\max}$ is the visual extinction for the sightline going through the IC 63 center.

\bibliography{ms.bbl}
\end{document}